\documentclass[oneside,graybox,sectrefs]{svmult}

\usepackage[utf8]{inputenc}
\usepackage{mathptmx}
\usepackage{helvet}
\usepackage{courier}
\usepackage{braket}
\usepackage{amsfonts}
\usepackage{amsmath}
\usepackage{type1cm}         
\usepackage{exercise}
\usepackage{graphicx}
\usepackage[bottom]{footmisc}
\usepackage[usenames,dvipsnames,x11names]{xcolor}
\usepackage{listings}
\usepackage{epic}
\usepackage{eepic}
\usepackage{a4wide}
\usepackage{color}
\usepackage{amsmath}
\usepackage{amssymb}
\usepackage[T1]{fontenc}
\usepackage{cite}
\usepackage{shadow}
\usepackage{hyperref}
\usepackage{bezier}
\usepackage{pstricks}
\usepackage{textcomp,type1ec,pdfpages}
\usepackage{bera}
\usepackage{slashed}
\usepackage{environ}
\usepackage{esvect}
\usepackage{xspace}
\usepackage{bm}

\newcommand{\ie}{\textit{i.e.}}
\newcommand{\eg}{\textit{e.g.}}

\newcommand{\cf}{\textit{cf.}\xspace}
\newcommand{\adhoc}{\textit{ad hoc}\xspace}
\newcommand{\etal}{\textit{et al.}\xspace}

\newcommand{\apriori}{\textit{a priori}\xspace}

\newcommand{\ii}{\mathrm{i}}
\newcommand{\eex}{\mathrm{e}}

\newcommand{\hc}{\mathrm{h.c.}}

\newcommand{\abs}[1]{\left|#1\right|}

\newcommand{\dd}{\mathrm{d}}

\newcommand{\vNabla}{\boldsymbol{\nabla}}
\newcommand{\Laplace}{\vNabla^2}
\newcommand{\dAlem}{\Box}

\newcommand{\vx}{\mathbf{x}}
\newcommand{\vy}{\mathbf{y}}

\newcommand{\vk}{\mathbf{k}}

\newcommand{\vq}{\mathbf{q}}

\newcommand{\vA}{\mathbf{A}}

\newcommand{\vZero}{\mathbf{0}}

\newcommand*\lrvec[1]{%
\ensuremath{\overset{\smash{\raisebox{-1.5pt}{\tiny$\leftrightarrow$}}}{#1}}%
}

\newcommand{\vNablaLR}{\lrvec{\vNabla}}

\newcommand{\skvec}[1]{\vv{#1}}

\newcommand{\vD}{\boldsymbol{D}}

\newcommand{\LL}{\mathcal{L}}

\newcommand{\eps}{\varepsilon}

\newcommand{\Mhi}{\ensuremath{M_{\rm hi}}}
\newcommand{\Mlo}{\ensuremath{M_{\rm lo}}}

\newcommand{\GeV}{\ensuremath{\mathrm{GeV}}}
\newcommand{\MeV}{\ensuremath{\mathrm{MeV}}}
\newcommand{\fm}{\ensuremath{\mathrm{fm}}}

\newcommand{\mpi}{m_\pi}
\newcommand{\msigma}{m_\sigma}
\newcommand{\mrho}{m_\rho}

\newcommand{\MN}{M_N}

\newcommand{\Md}{M_d}
\newcommand{\Bd}{B_d}

\newcommand{\fpi}{f_\pi}

\newcommand{\gamt}{\gamma_t}
\newcommand{\at}{a_t}

\newcommand{\rhot}{\rho_t}

\newcommand{\gamd}{\gamt}

\NewEnviron{skmattwod}{%
\left(\begin{array}{cc}
  \BODY
\end{array}\right)
}

\NewEnviron{skvectwod}{%
\left(\begin{array}{c}
  \BODY
\end{array}\right)
}

\newcommand{\twodmat}[1]{\begin{skmattwod}#1\end{skmattwod}}
\newcommand{\twodvec}[1]{\begin{skvectwod}#1\end{skvectwod}}

\newcommand{\diag}{\mathrm{diag}}

\newcommand{\one}{\mathbf{1}}
\newcommand{\leviciv}{\epsilon}

\newcommand{\Tr}{\mathrm{Tr}}

\newcommand{\idx}[2]{{}^{#1}{}_{#2}}
\newcommand{\idxx}[3]{\big(#1\big)\idx{#2}{#3}}
\newcommand{\idxxx}[3]{\left(#1\right)^{\!\!#2}_{\,#3}}

\newcommand{\mathspace}{\ \ }
\newcommand{\mathtext}[1]{\mathspace\text{#1}\mathspace}

\newcommand{\OO}{\mathcal{O}}

\newcommand{\commbig}[2]{\big[#1,#2\big]}

\newcommand{\acommbig}[2]{\big\{#1,#2\big\}}

\NewEnviron{subalign}[1][]{%
\begin{subequations}\begin{align}
  \BODY
\end{align}\label{#1}\end{subequations}
}

\NewEnviron{spliteq}{%
\begin{equation}\begin{split}
  \BODY
\end{split}\end{equation}
}

\newcommand*{\vcenteredhbox}[1]
{\begingroup\setbox0=\hbox{#1}\parbox{\wd0}{\box0}\endgroup}

\newcommand{\ThreeSOne}{\ensuremath{{}^3S_1}\xspace}
\newcommand{\OneSNot}{\ensuremath{{}^1S_0}\xspace}

\newcommand{\Triton}{\ensuremath{{}^3\mathrm{H}}\xspace}
\newcommand{\ThreeH}{\Triton}

\newcommand{\epsgammaout}{\vec{\epsilon}^{\;*}_{s_\gamma}}

\newcommand{\epsdout}{\vec{\epsilon}^{\;*}_{s_d}}

\newcommand{\y}{y}
\newcommand{\yt}{\ensuremath{\y_t}}
\newcommand{\ys}{\ensuremath{\y_s}}

\newcommand{\sigt}{\ensuremath{g_t}}
\newcommand{\sigs}{\ensuremath{g_s}}

\newcommand{\st}{\ensuremath{{s,t}}}
\newcommand{\yst}{\ensuremath{\y_\st}}
\newcommand{\sigst}{\ensuremath{g_\st}}

\title{General aspects of effective field theories and few-body applications}

\author{Hans-Werner Hammer and Sebastian K\"onig}
\institute{Hans-Werner Hammer
\at Institut f\"ur Kernphysik,
Technische Universit\"at Darmstadt,
64289 Darmstadt, 
Germany,
\email{Hans-Werner.Hammer@physik.tu-darmstadt.de},
\and Sebastian K\"onig
\at Department of Physics,
The Ohio State University,
Columbus, Ohio 43210,
USA
\email{koenig.389@osu.edu}
}

\graphicspath{{Chapter4-figures/}}

\makeatletter
\ds@numart
\makeatother

\begin{document}

\maketitle
\abstract{Effective field theory provides a powerful framework to 
exploit a separation of scales in physical systems. In these lectures,
we discuss some general aspects of effective field theories and their 
application to few-body physics. 
In particular, we consider an effective field theory for 
non-relativistic particles with resonant short-range interactions
where certain parts of the interaction need to be treated nonperturbatively.
As an application, we discuss the so-called \emph{pionless effective field
theory} for low-energy nuclear physics. The extension to include long-range 
interactions mediated by photon and pion-exchange is also addressed.}

\section{Introduction: dimensional analysis and the separation of scales}
\label{sec:EFT-Intro}

Effective field theory (EFT) provides a general approach to calculate low-energy 
observables by exploiting scale separation.  The origin of the EFT approach can 
be traced to the development of the renormalization group~\cite{Wilson-83} and 
the intuitive understanding of ultraviolet divergences in quantum field 
theory~\cite{Lepage-89}.  A concise formulation of the underlying principle was 
given by Weinberg~\cite{Weinberg:1978kz}: If one 
starts from the most general Lagrangian consistent with the symmetries of the 
underlying theory, one will get the most general S-matrix consistent with these 
symmetries.  As a rule, such a most general Lagrangian will contain infinitely 
many terms.  Only together with a power counting scheme that orders these terms 
according to their importance at low energies one obtains a predictive paradigm 
for a low-energy theory.

The Lagrangian and physical observables are typically 
expanded in powers of a low-momentum scale $\Mlo$,
which can be a typical external momentum or an internal
infrared scale, over a high-momentum scale 
$\Mhi\gg\Mlo$.\footnote{Note there are often 
more than two scales, which complicates the power counting. Here we focus on
the simplest case to introduce the general principle.}
This expansion provides the basis for the power counting scheme.
It depends on the system to which physical scales 
$\Mhi$ and $\Mlo$ correspond to.  

\begin{figure}[tb]
\centerline{\includegraphics*[width=8cm]{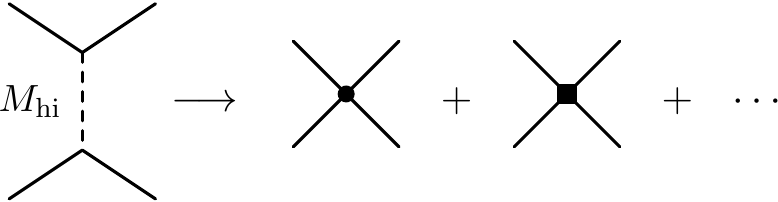}}
\caption{Expansion of heavy-particle exchange
between light particles in terms of contact interactions between
light particles. The solid and dashed lines denote light and heavy
particles, respectively. The circle and square
denote contact interactions with zero and two derivatives, in order.
\label{fig:resosat}}
\end{figure}

As an example, we take a theory that is made of two
particle species, two light bosons with mass $\Mlo$ and heavy
bosons with mass $\Mhi \gg \Mlo$.\footnote{For further examples,
see the lectures by Kaplan \cite{Kaplan:1995uv,Kaplan:2005es}.}
We consider now soft processes
in which the energies and momenta are of the order of the 
light particle mass (the so-called soft scale). 
Under such conditions, the short-distance
physics related to the heavy particles can never be resolved explicitly.
However, it can be represented by light-particle contact interactions 
with increasing dimension (number of derivatives). To illustrate this,
we consider the scattering of the light particles mediated by
heavy-particle exchange, with
$g$ the heavy-light coupling constant. The corresponding interaction
Lagrangian is given by
\begin{equation}
 {\mathcal L}_{\rm int}=g\left(\chi^\dagger \phi \phi +
 \phi^\dagger \phi^\dagger \chi\right)\,,
\end{equation}
where $\phi$ denotes the light boson field and
$\chi$ is the heavy boson field. As depicted in
Fig.~\ref{fig:resosat}, one can represent such exchange diagrams by a 
sum of local operators of the light fields with increasing
number of derivatives. In a symbolic notation, the leading order 
scattering amplitude can be written as
\begin{equation}
 T\sim \frac{g^2}{\Mhi^2 - q^2} = \frac{g^2}{\Mhi^2} + 
 \frac{g^2 \, q^2}{\Mhi^4} + \cdots \,,
\end{equation}
with $q^2$ the squared 4-momentum transfer.  We will come back
to this example in more detail in section~\ref{sec:EFT-Basics}.

In many cases, the corresponding high-energy theory
is either not known or can not easily be solved. 
Still, EFT offers a predictive
and systematic framework for performing calculations in the
light-particle sector. We denote by $Q$ a typical energy
or momentum of the order of $\Mlo$ and by $\Mhi$ the
hard scale where the EFT will break down. In many cases, this
scale is set by the masses of the heavy particles not considered
explicitly and thus replaced by contact interactions as in the example above.  
In such a setting, any matrix element or Green's
function admits an expansion in the small parameter 
$Q/\Mhi$~\cite{Weinberg:1978kz}
\begin{equation}
 {\mathcal M} = \sum\limits_{\nu} \left(\frac{Q}{\Mhi}\right)^{\nu}  
 \,{\mathcal F} \left(\frac{Q}{\Lambda}, g_i\right)
\end{equation}
where ${\mathcal F}$ is a function of order one (this is the naturalness
assumption), $\Lambda$
a regularization scale (related to the UV divergences appearing
in the loop graphs) and the $g_i$ denotes a collection of
coupling constants, often called low-energy constants (LECs).
These parameterize (encode) the unknown high-energy (short-distance)
physics and must be determined by a fit to data (or
can be directly calculated if the corresponding high-energy theory
is known/can be solved). The counting index $\nu$ in general depends 
on the fields in the effective theory, the number of derivatives and
the number of loops. This defines the so-called power counting 
which allows to categorize all contributions to any matrix element
at a given order. It is important to stress that $\nu$ must be
bounded from below to define a sensible EFT. In QCD, \eg, this is 
a consequence of the spontaneous breaking of chiral symmetry. 

The contributions with the lowest possible value of $\nu$ 
define the so-called leading order (LO) contribution, the first
corrections with the second smallest allowed value of $\nu$ the
next-to-leading order (NLO) terms and so on. In contrast to more
conventional perturbation theory, the small parameter is not a 
dimensionless coupling constant (like, \eg, in Quantum Electrodynamics) 
but rather a ratio of two scales. Typically, one
expands in the ratio of a small energy or momentum and 
the hard scale $\Mhi$. A prototype of such
a perturbative EFT is chiral perturbation theory that exploits the
strictures of the spontaneous and explicit chiral symmetry breaking
in QCD~\cite{Gasser:1983yg,Gasser:1984gg}.  Here, the light degrees
of freedom are the pions, that are generated through the symmetry violation.
Heavier particles like \eg vector mesons only appear indirectly
as they generate local four-pion interactions with four, six, etc
derivatives.

In these lectures, we also consider EFTs with bound states, where 
certain contributions need to resummed nonperturbatively. In
section~\ref{sec:EFT-Basics}, we start with some general considerations.
This is followed by the explicit discussion of an EFT for 
non-relativistic bosons with short-range interactions 
and large scattering length in section~\ref{sec:EFT-Bosons}.  The extension of 
this framework 
to low-energy nucleons is presented in section~\ref{sec:EFT-Nucleons}
Finally, we will discuss the inclusion of long-range interactions
mediated by photon and pion exchange in~\ref{sec:EFT-Beyond}.

\section{Theoretical foundations of effective field theory}
\label{sec:EFT-Basics}

As mentioned in the introduction, effective field theories are described by 
writing down Lagrangians with an infinite number of terms, restricted only by 
symmetry considerations, and ordered by a scheme referred to as ``power 
counting.''  In this section, we discuss the meaning and importance of all 
these ingredients.

\subsection{Top-down vs.\ bottom-up approaches}

Generally, there are two different motivations for working with an EFT.  Given a 
known quantum field theory, which can be solved to compute a given quantity of 
interest, it can be beneficial to switch to an effective description valid only 
in a limited energy regime simply because carrying out the calculation is more 
efficient with the effective theory.  With such a solvable underlying theory, 
the parameters (``low-energy constants'') of the effective theory can be 
computed directly by considering some number of (simple) processes, \ie, one 
does not need experimental input beyond what was needed to fix the parameters 
of the underlying theory.  This approach, based on a reduction of expressions 
from the underlying to the effective picture is called a ``top-down'' approach.

An alternative procedure, somewhat closer to what we described at the outset, 
is to start ``bottom up,'' \ie, by simply writing down the effective 
Lagrangian directly---or more precisely only those terms of the infinitely 
many which are needed to achieve a given desired accuracy.  Being able to 
do that of course requires that as a first step one has already figured out 
which terms are allowed and how they should be ordered.

Our approach here is to work top down in the pedagogical sense, \ie, postpone 
the discussion of the bottom-up approach and its ingredients until later in 
this section, and instead dive into the matter starting with examples that show 
how effective low-energy theories can arise from more fundamental ones.  We 
assume that the reader is familiar the material from a standard 
(relativistic) quantum field theory course.

\subsubsection{Integrating out exchange particles: part I}
\label{sec:EFT-IntOut-1}

As was also mentioned in the introduction, the very first step in 
the construction of an EFT is to identify the relevant degrees of freedom to 
work with, as well as those which are irrelevant and thus do not need to 
be kept explicitly (with emphasis on the last word, because 
\emph{implicitly} the physics of left-out degrees of freedom should and does 
enter in the effective description).

\begin{figure}[htbp]
\centering
\includegraphics[clip,width=0.75\textwidth]{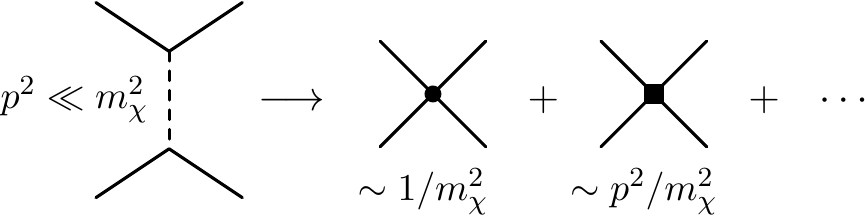}
\caption{Chain of contact interactions obtained by integrating out an exchange 
particle.}
\label{fig:BosonContact-2body}
\end{figure}

Let us illustrate this by showing how integrating out a ``heavy'' 
particle gives rise to contact interactions between the remaining degrees 
of freedom (see the example in section \ref{sec:EFT-Intro}).  We 
stress that the particles which are integrated out can in fact be lighter than 
what is left (like it is the case in pionless EFT)---what really matters for 
the procedure is which particles are assumed to appear in \emph{asymptotic} 
states, and what is the typical energy/momentum scale between those.  In that 
spirit, we are not making explicit assumptions about the mass hierarchy of the 
particles in the following.  For the illustration here, we consider two scalar 
fields (complex and relativistic) with Yukawa interactions and start with a 
Lagrangian for two species:
\begin{equation}
 \LL = {-}\phi^\dagger\left(\dAlem + m_\phi^2\right)\phi
 - \chi^\dagger\left(\dAlem + m_\chi^2\right)\chi
 + g\left(\phi^\dagger\phi^\dagger\chi + \hc\right) \,.
\label{eq:L-phi-chi-rel}
\end{equation}
Suppose now we are only interested in interactions between $\phi$ particles 
at energy scales much smaller than $m_\chi$, so that the explicit 
$\chi$ exchange generated by the interaction term in 
Eq.~\eqref{eq:L-phi-chi-rel} cannot be resolved.  In that case, we can derive a 
new effective Lagrangian that only contains $\phi$ degrees of freedom, a 
process referred to as ``integrating out'' the field $\chi$ stemming from its 
implementation in the path-integral formalism.  In effect, that amounts to 
using the equations of motion, which we do here.  From the Euler-Lagrange 
equation for $\chi^\dagger$, we directly get
\begin{equation}
 \chi = \left(\dAlem + m_\chi^2\right)^{-1} g\phi\phi \,.
\label{eq:chi-g-phiphi}
\end{equation}
Defining the Klein-Gordon propagator
\begin{equation}
 D_\chi(x-y)
 = \int\frac{\dd^4p}{(2\pi)^4} \eex^{{-}\ii p(x-y)}
 \frac{\ii}{p^2-m_\chi^2 + \ii\eps} \,,
\label{eq:D-chi-KG}
\end{equation}
satisfying
\begin{equation}
 \left(\dAlem + m_\chi^2\right)D_\chi(x-y) = {-}\ii \delta^{(4)}(x-y) \,,
\label{eq:chi-GF}
\end{equation}
we can write out Eq.~\eqref{eq:chi-g-phiphi} in configuration space as
\begin{equation}
 \chi(x) = \ii g \int\dd^4y \, D_\chi(x-y) \phi(y) \phi(y) \,.
\end{equation}
Inserting this back into the Lagrangian~\eqref{eq:L-phi-chi-rel}, we obtain 
\begin{equation}
 \LL(x) = {-}\phi^\dagger(x)\left(\dAlem + m_\phi^2\right)\phi(x)
 - \ii g^2 \phi^\dagger(x) \phi^\dagger(x)
 \int\dd^4y \, D_\chi(x-y) \phi(y) \phi(y) \,,
\label{eq:L-phi-chi-rel-nl}
\end{equation}
where we have written out the spacetime dependence of all fields and used 
Eq.~\eqref{eq:chi-GF} to cancel the terms involving $\chi^\dagger(x)$.  So far, 
we have made only exact manipulations, but the resulting 
Lagrangian~\eqref{eq:L-phi-chi-rel-nl} is \emph{non-local}, \ie, it depends on 
fields evaluated at different spacetime points.  To simplify it further, we 
want to exploit the fact that $\chi$ is considered ``heavy'' compared to the 
scales we want to describe.  Mathematically, this means that $D_\chi(x-y)$ is 
peaked at distances that are small compared to $1/m_\chi^2$.  There are several 
ways to implement this knowledge.  A particularly intuitive version is to expand 
the propagator~\eqref{eq:D-chi-KG} in momentum space,
\begin{equation}
 \frac{\ii}{p^2-m_\chi^2 + \ii\eps}
 = \frac{{-}\ii}{m_\chi^2}\left(1 + \frac{p^2}{m_\chi^2} + \cdots\right) \,,
\label{eq:D-chi-expansion}
\end{equation}
and then Fourier-transform back to configuration space.  The first term gives a 
simple delta function, and terms with powers of $p^2$ induce operators with 
derivatives acting on $\delta(x-y)$.  Inserting the leading term into 
Eq.~\eqref{eq:L-phi-chi-rel-nl}, we arrive at the effective \emph{local} 
Lagrangian
\begin{equation}
 \LL_{\text{eff}}(x) = {-}\phi^\dagger(x)\left(\dAlem + m_\phi^2\right)\phi(x)
 - \frac{g^2}{m_\chi^2} \phi^\dagger(x) \phi^\dagger(x) \, \phi(x) \, \phi(x)
 + \cdots \,.
\label{eq:L-phi-chi-rel-eff}
\end{equation}
The ellipses contain operators with derivatives acting on $\phi(x)$, obtained 
from those acting on the delta functions from the propagator after integrating 
by parts.  A diagrammatic illustration of the procedure is shown in
Fig.~\ref{fig:BosonContact-2body}.

We note that an alternative derivation of the above 
result, discussed for example in Ref.~\cite{Donoghue:1992}, is given by 
Taylor-expanding the field product $\phi(y) \phi(y)$ about $y=x$ under the 
integral and then using the properties of the propagator.  This directly gives 
terms with an increasing number of derivatives acting on $\phi^\dagger(x) 
\phi^\dagger(x)$, and those with an odd number of derivatives are found to 
vanish, in agreement with Eq.~\eqref{eq:D-chi-expansion} featuring only even 
powers of $p^2$.

\subsubsection{Emergence of many-body forces}

We now add a third field $\Phi$ to the Lagrangian:
\begin{multline}
 \LL = {-}\phi^\dagger\left(\dAlem + m_\phi^2\right)\phi
 - \chi^\dagger\left(\dAlem + m_\chi^2\right)\chi
 - \Phi^\dagger\left(\dAlem + m_\Phi^2\right)\Phi \\
 \null + g\left(\phi^\dagger\phi^\dagger\chi + \hc\right)
 + g'\left(\Phi^\dagger\phi\chi + \hc\right) \,.
\label{eq:L-phi-chi-Phi-rel}
\end{multline}
The new interaction term is chosen such that $\Phi$ can ``decay'' into a $\phi$ 
and a $\chi$, thus acting like a heavier version of the $\phi$.  In spite of 
the simplicity of this bosonic toy model, it is useful to think about $\phi$ 
and $\Phi$ as the nucleon and its $\Delta$ excitation, respectively, and about 
$\chi$ as a pion field.
If we first integrate out the $\Phi$ field following the procedure described in 
the previous section, we find
\begin{equation}
 \Phi(x) = \ii g' \int\dd^4y \, D_\Phi(x-y) \phi(y) \chi(y) \,,
\end{equation}
and thus
\begin{equation}
 \LL_{\text{eff}} = {-}\phi^\dagger\left(\dAlem + m_\phi^2\right)\phi
 - \chi^\dagger\left(\dAlem + m_\chi^2\right)\chi
 + g\left(\phi^\dagger\phi^\dagger\chi + \hc\right)
 + \frac{g'^2}{m_\Phi^2} \phi^\dagger\chi^\dagger\phi\chi
 + \cdots \,,
\label{eq:L-Phi-phi-chi-rel-eff-1}
\end{equation}
where we have only kept the leading (no derivatives) induced contact 
interaction.  Proceeding as before for the $\chi$ field, we now get
\begin{equation}
 \left(\dAlem + m_\chi^2\right)\chi = g\phi\phi
 + \frac{g'^2}{m_\Phi^2} \phi^\dagger \phi \, \chi 
 + \cdots \,.
\label{eq:chi-g-phiphi-etc-1}
\end{equation}
This can no longer be solved exactly because we now have a $\chi$ on the 
right-hand side.  However, using the general operator identify
\begin{equation}
 \left(\hat{A}-\hat{B}\right)^{-1}
 = \hat{A}^{-1} + \hat{A}^{-1}\hat{B}\,\hat{A}^{-1} + \cdots \,,
\label{eq:AB-inv}
\end{equation}
we can write down a formal iterative solution:
\begin{equation}
 \chi = \left(\dAlem + m_\chi^2\right)^{-1} g\phi\phi
 + \left(\dAlem + m_\chi^2\right)^{-1}\frac{g'^2}{m_\Phi^2}\phi^\dagger\phi
 \left(\dAlem + m_\chi^2\right)^{-1} g\phi\phi + \cdots \,,
\label{eq:chi-g-phiphi-etc-2}
\end{equation}
with each of the inverse differential operators giving a propagator when 
written out.  Those, in turn, each give factors of ${-}\ii/m_\Phi^2$ times a 
delta function, plus additional terms with derivatives.

\begin{prob}
{\emph Exercise:} Derive Eq.~\eqref{eq:AB-inv}.
\end{prob}

\begin{figure}[htbp]
\centering
\includegraphics[clip,width=0.85\textwidth]{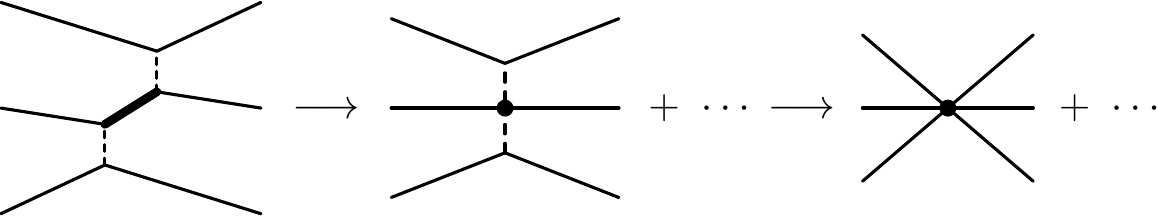}
\caption{Emergence of a three-body contact interaction.}
\label{fig:BosonContact-3body}
\end{figure}

Inserting the above result back into the Eq.~\eqref{eq:L-Phi-phi-chi-rel-eff-1}, 
we see that in addition to the two-body contact operator $(\phi^\dagger\phi)^2$ 
obtained previously, we now also get all kinds of higher-body interactions.  For 
example, we get a three-body force through
\begin{equation}
 \frac{g'^2}{m_\Phi^2} \phi^\dagger\chi^\dagger\phi\chi
 \rightarrow \frac{g'^2 g^2}{m_\Phi^2 m_\chi^4} (\phi^\dagger\phi)^3 \,.
\end{equation}
In Fig.~\ref{fig:BosonContact-3body} it is illustrated diagrammatically how 
such a term arises subsequently, starting from a diagram derived from the 
original Lagrangian~\eqref{eq:L-phi-chi-Phi-rel} with three fields.

\subsection{Nonrelativistic field theory}

Relativistic effects and exact Lorentz invariance are not crucial to describe 
systems at low energies, where ``low'' means ``much smaller than the particles' 
rest mass.''  Based on that, one typically starts with a nonrelativistic 
framework and writes down effective Lagrangians of so-called Schr\"odinger 
fields, \eg,
\begin{equation}
 \LL_{\phi,\text{free}}
 = \phi^\dagger \left(\ii\partial_t + \frac{\Laplace}{2m}\right)\phi
\label{eq:L-phi-free-nonrel}
\end{equation}
for a free scalar particle, where $\phi^\dagger(t,\vx)$ is the field operator 
that creates a particle at time $t$ and position $\vx$, and $\phi(t,\vx)$)
correspondingly destroys it.  Written in terms of momentum-space ladder 
operators $\hat{a}_{\mathbf{p}}$,$\hat{a}^\dagger_{\mathbf{p}}$ (as they appear in standard 
many-body quantum mechanics), we have
\begin{equation}
 \phi(t,\vx)
 = \int\!\frac{\dd^3p}{(2\pi)^3} \hat{a}_{\mathbf{p}} \,\eex^{-\ii E_{\mathbf{p}} t} 
 \eex^{\ii{\mathbf{p}}\cdot\vx} \,,
\end{equation}
and analogously for $\phi^\dagger(t,\vx)$.  Note that here $E_{\mathbf{p}} = 
{\mathbf{p}}^2/(2m)$ is the kinetic energy alone, and that creation and destruction 
operators are completely separated.  Intuitively, this makes perfect sense: 
At low energies, virtual particle-antiparticle pairs would be highly 
off-shell, thus giving rise to very short-range effects that we can 
simply describe as contact interactions.  Other effects, such as self-energy 
corrections to the particle mass, are automatically accounted for by using the 
physical value for $m$ in Eq.~\eqref{eq:L-phi-free-nonrel}.  With this in mind, 
one can proceed in the bottom-up approach and construct an interacting theory 
by supplementing the free Lagrangian with all allowed contact operators.  In 
particular, like Eq.~\eqref{eq:L-phi-free-nonrel} they should all be invariant 
under Galilei transformations, the low-energy remnant of the Poincar{\'e} 
group.  Before we come back to this, however, we find it instructive to 
explicitly consider the low-energy limit of a relativistic theory.

\subsubsection{Nonrelativistic limit of a bosonic field}
\label{sec:EFT-NonRelBos}

Let us make the connection of Eq.~\eqref{eq:L-phi-free-nonrel} to a
relativistic complex Klein--Gordon field $\Phi$, the Lagrangian for which can be 
written as
\begin{equation}
 \LL_{\varphi,\text{free}}
 = {-}\varphi^\dagger\left(\partial_t^2 - \Laplace + m^2\right)\varphi 
\,.
\end{equation}
Using integration by parts, this can be shown to be equivalent to the more 
common form written with $(\partial_\mu\varphi^\dagger)(\partial^\mu\varphi)$.  
This implies the Klein--Gordon equation for the field operator,
\begin{equation}
 \left(\partial_t^2 - \Laplace + m^2\right)\varphi = 0 \,,
\end{equation}
the most general solution of which is typically written as (with a 
four-vectors $x=(t,\vx)$, $p=(p_0,{\mathbf{p}})$, and a Lorentz-invariant integration 
measure)
\begin{equation}
 \varphi(x)
 = \int\!\frac{\dd^3p}{(2\pi)^3} \frac{1}{\sqrt{2\omega_{\mathbf{p}}}}\left(
 \hat{a}_{\mathbf{p}}\,\eex^{-\ii p\cdot x} + \hat{b}^\dagger_{\mathbf{p}}\,\eex^{\ii p\cdot x}
 \right)\Bigg|_{p_0=\omega_{\mathbf{p}}} \,,
\label{eq:Phi}
\end{equation}
where $\omega_{\mathbf{p}} = \sqrt{{\mathbf{p}}^2+m^2}$.  With this convention where $p_0$ is 
chosen positive, modes created by $\hat{a}^\dagger_{\mathbf{p}}$ correspond to particles 
(propagating forward in time), whereas $\hat{b}^\dagger_{\mathbf{p}}$ creates an 
antiparticle (positive-energy state propagating backwards in time).  That we 
have both stems from the fact that the complex scalar field corresponds to two 
real ones (completely decoupled in the absence of interactions), each of which 
comes with its own pair of creation and annihilation operators.  To take 
the nonrelativistic limit, we have to consider the particle and antiparticles 
separately.  Defining
\begin{equation}
 \varphi_a(x) = \int\!\frac{\dd^3p}{(2\pi)^3} \frac{1}{\sqrt{2\omega_{\mathbf{p}}}}
 \hat{a}_{\mathbf{p}}\,\eex^{-\ii p\cdot x}\Bigg|_{p_0=\omega_{\mathbf{p}}}
 \equiv \eex^{-\ii m t} \phi_a(x) \,,
\label{eq:Phi-phi-a}
\end{equation}
and plugging this into the Klein--Gordon equation, we get
\begin{equation}
 \eex^{-\ii m t} \left[\partial_t^2 - 2\ii m\,\partial_t
 - \vNabla^2\right] \phi_a(x) = 0 \,,
\label{eq:KG-phi-a}
\end{equation}
where the quadratic mass term has canceled.  Since $\phi_a(x) = \eex^{\ii m t} 
\varphi_a(x)$, we see from Eq.~\eqref{eq:Phi-phi-a} that in the Fourier 
transform each time derivative acting on $\phi_a(x)$ brings down a factor
\begin{equation}
 \omega_p - m = \sqrt{{\mathbf{p}}^2 + m^2} - m \approx {\mathbf{p}}^2/(2m) \,,
\end{equation}
\ie, just the kinetic energy $E_{\mathbf{p}}$ up to corrections of higher order in 
$1/m$.  In the nonrelativistic limit, $E_{\mathbf{p}} \ll m$, so we see that we can 
neglect the quadratic time derivative in Eq.~\eqref{eq:KG-phi-a} compared to the 
other two terms in Eq.~\eqref{eq:KG-phi-a}, and then recover the Schr\"odinger 
equation for $\phi_a$:
\begin{equation}
 \left(\ii\partial_t + \frac{\vNabla^2}{2m}\right)\phi_a(x) = 0 \,.
\end{equation}
This establishes the connection to our $\phi(t,\vx)$ in 
Eq.~\eqref{eq:L-phi-free-nonrel} when we insert an additional factor 
$\sqrt{2m}$ in the field redefinition to account for the otherwise different 
normalizations.  For the antiparticles, we can carry out an analogous procedure, 
except that we have to choose the opposite sign for the mass-dependent phase in 
the field redefinition analogous to Eq.~\eqref{eq:Phi-phi-a} because the 
antiparticle part of $\varphi(x)$ comes with a factor $\eex^{+\ii p\cdot x}$.

\subsubsection{Nonrelativistic fermions}
\label{sec:EFT-NonRelFerm}

For relativistic Dirac fermions, the nonrelativistic reduction can be carried 
out with the help of a so-called Foldy-Wouthuysen transformation.\footnote{An 
alternative way to perform the nonrelativistic reduction is to introduce a 
``heavy fermion'' field~\cite{Jenkins:1990jv}.  A comparison of this formalism 
and the Foldy-Wouthuysen transformation can be found in 
Ref.~\cite{Gardestig:2007mk}.}  The idea 
behind the approach is to decouple the particle and antiparticle modes 
contained together in a four-spinor $\psi$ through a sequence of unitary 
transformations.  In the following, we demonstrate this procedure, using an 
\emph{interacting} model theory to also illustrate what happens to interaction 
terms in the nonrelativistic limit.  Since it will be useful to motivate the 
pionless EFT discussed in Sec.~\ref{sec:EFT-Nucleons}, we start with a 
Lagrangian of the form
\begin{multline}
 \LL = \bar{\psi}\left(\ii\slashed{\partial} - \MN\right)\psi
 + \frac12(\partial^\mu\skvec{\pi}) \cdot (\partial_\mu\skvec{\pi})
 - \frac12\mpi^2 \skvec{\pi}^2
 + \frac12(\partial^\mu\sigma) \cdot (\partial_\mu\sigma)
 - \frac12\msigma^2 \sigma^2 \\
 \null - g\bar{\psi}(\sigma - \ii \gamma^5\skvec{\tau}\cdot\skvec{\pi})\psi \,,
\label{eq:L-pi-N-PS}
\end{multline}
where the nucleon field $\psi$ is an isospin doublet of Dirac spinors, 
$\vec{\pi}$ is an isospin triplet, and $\sigma$ is an isoscalar.  A Lagrangian 
of this form (plus additional interaction terms among $\sigma$ and 
$\vec{\pi}$), can be obtained from a linear sigma model after spontaneous 
symmetry breaking (see, for example, \cite{Donoghue:1992}, Chapter I) and 
augmented by an explicit mass term for $\vec{\pi}$.\footnote{We stress, 
however, that this really is a model and not a proper EFT describing QCD.}  We 
denote the Pauli matrices in spin and isospin space as $\vec{\sigma} = 
(\sigma^i)$ and $\vec{\tau} = (\tau^\lambda)$, respectively.  For the gamma 
matrices we use the standard (Dirac) representation:
\begin{equation}
 \gamma^0 = \twodmat{\one & 0 \\ 0 & {-}\one} \mathtext{,}
 \gamma^i = \twodmat{0 & \sigma^i \\ {-}\sigma^i & 0} \mathtext{,}
 \gamma^5 = \twodmat{0 & \one \\ \one & 0} \,.
\end{equation}

To perform the nonrelativistic reduction, we start by separating odd and even 
operators, which are two-by-two block matrices in Dirac space.  The result is
\begin{equation}
 \LL_\psi = \psi^\dagger\left(\hat{E} + \hat{O} - \gamma^0\MN\right)\psi \,,
\label{eq:L-E-O}
\end{equation}
where
\begin{equation}
 \hat{E} = \twodmat{
  \ii\partial_t - g\sigma & 0 \\
  0 & \ii\partial_t + g\sigma
 }
 \mathtext{and}
 \hat{O} = \twodmat{
   0 & {-}\ii\skvec{\sigma}\cdot\vNabla + \ii g\skvec{\tau}\cdot\skvec{\pi} \\
   {-}\ii\skvec{\sigma}\cdot\vNabla - \ii g\skvec{\tau}\cdot\skvec{\pi} & 0
 } \,.
\end{equation}
Rotating the phase of the fermion field,
\begin{equation}
 \psi \rightarrow \tilde{\psi} = \eex^{{-}\ii\MN t}\psi \,,
\end{equation}
just like we did for the bosonic field in Eq.~\eqref{eq:Phi-phi-a}, we can 
remove the mass term for the upper components:
\begin{equation}
 \LL_\psi
 = \tilde{\psi}^\dagger\left(\hat{E} + \hat{O}
 - (\gamma^0-\one)\MN\right)\tilde{\psi} \,.
\label{eq:L-E-O-tilde}
\end{equation}
The Foldy-Wouthuysen transformation is now constructed to (approximately) 
decouple the upper from the lower components, \ie, nucleons from their 
antiparticles.  To achieve this, we use a sequence of further unitary 
redefinitions of the fermion field.  The first of these is
\begin{equation}
 \tilde\psi \rightarrow \tilde{\psi}' = \eex^{{-}\ii\hat{S}}\tilde{\psi}
 \mathtext{with}
 \hat{S} = {-}\frac{\ii\gamma^0\hat{O}}{2\MN} \,.
\end{equation}
Let us consider this transformation up to quadratic order in $1/\MN$.  
Expanding the exponential, we have
\begin{equation}
 \tilde{\psi} = \eex^{\ii\hat{S}} \tilde{\psi}'
 = \left(1 + \frac{\gamma^0\hat{O}}{2\MN}
 + \frac{\big(\gamma^0\hat{O}\big)^2}{8\MN^2}
 + \OO\big(1/\MN^3\big)\right) \tilde{\psi}'
\end{equation}
and likewise
\begin{equation}
 \tilde{\psi}^\dagger = \tilde{\psi}'^\dagger\eex^{{-}\ii\hat{S}} 
 = \tilde{\psi}'^\dagger \left(1 - \frac{\gamma^0\hat{O}}{2\MN}
 + \frac{\big(\gamma^0\hat{O}\big)^2}{8\MN^2}
 + \OO\big(1/\MN^3\big)\right) \,.
\end{equation}
Inserting this into Eq.~\eqref{eq:L-E-O-tilde} and collecting contributions up 
to corrections which are $\OO(1/\MN^2)$, we get a number of terms:
\begin{subequations}
\begin{equation}
 {-}\frac{\gamma^0\hat{O}\hat{E}}{2\MN} + \frac{\hat{E}\gamma^0\hat{O}}{2\MN}
 = \frac{\gamma^0\commbig{\hat{O}}{\hat{E}}}{2\MN} \,
\end{equation}
\begin{equation}
 \frac{\hat{O}\gamma^0\hat{O}}{2\MN}
 - (\gamma^0-\one)\frac{\big(\gamma^0\hat{O}\big)^2}{8\MN}
 - \frac{\gamma^0\hat{O}^2}{2\MN}
 + \frac{\gamma^0\hat{O}}{2\MN}(\gamma^0-\one)\frac{\gamma^0\hat{O}}{2\MN}
 - \frac{\big(\gamma^0\hat{O}\big)^2}{8\MN}(\gamma^0-\one)
 = {-}\frac{\gamma^0\hat{O}^2}{2\MN} \,,
\end{equation}
\begin{equation}
 \frac12\gamma^0\hat{O}(\gamma^0-\one) - \frac12(\gamma^0-\one)\gamma^0\hat{O}
 = -\hat{O} \,.
\end{equation}
\end{subequations}
Above we have used that
\begin{equation}
 \commbig{\gamma_0}{\hat{E}} = 0
 \mathtext{,}
 \acommbig{\gamma_0}{\hat{O}} = 0 \,,
\end{equation}
and $(\gamma^0)^2=\one$.  Collecting everything, we get
\begin{equation}
 \LL_{\psi}
 = \tilde{\psi}'^\dagger\left(
 \hat{E} - \frac{\gamma^0\hat{O}^2}{2\MN}
 + \frac{\gamma^0\commbig{\hat{O}}{\hat{E}}}{2\MN}
 - (\gamma^0-\one)\MN\right)\tilde{\psi}'
 + \OO(1/\MN^2) \,.
\label{eq:L-E-O-tilde-prime}
\end{equation}
The $\hat{O}^2$ term is even and we see that the original odd operator is 
canceled, but we have generated a new term $\sim\commbig{\hat{O}}{\hat{E}}$ 
If we neglect the interaction and consider
\begin{equation}
 \hat{E} = \hat{E}_{\text{free}} = \twodmat{\ii\partial_t & 0 \\ 0 & 
 \ii\partial_t}
 \mathtext{and}
 \hat{O} = \hat{O}_{\text{free}} = \twodmat{
   0 & {-}\ii\vec{\sigma}\cdot\vNabla \\
   {-}\ii\vec{\sigma}\cdot\vNabla & 0
 } \,,
\end{equation}
we find that $\commbig{\hat{O}_{\text{free}}}{\hat{E}_{\text{free}}}=0$ (partial 
derivatives commute) and thus the desired decoupling up to $\OO(1/\MN^2)$.  For 
the interacting case, on the other hand, the commutator does not vanish.  We 
see, however, that the new odd contribution is suppressed by a factor $1/\MN$.  
To push it to the next higher order, we need another rotation:
\begin{subequations}
\begin{equation}
 \tilde\psi' \rightarrow \tilde{\psi}'' = \eex^{{-}\ii\hat{S'}}\tilde{\psi} \,,
\end{equation}
with
\begin{equation}
 \hat{S'} = {-}\frac{\ii\gamma^0\hat{O}'}{2\MN}
 \mathtext{with}
 \hat{O}' = \frac{\gamma^0\commbig{\hat{O}}{\hat{E}}}{2\MN}
\end{equation}
\label{eq:S-prime}
\end{subequations}
After a couple of steps, we arrive at
\begin{equation}
 \LL_\psi
 = \tilde{\psi}''^\dagger\left(
 \hat{E} - \frac{\gamma^0\hat{O}^2}{2\MN}
 - (\gamma^0-\one)\MN\right)\tilde{\psi}''
 + \OO(1/\MN^2) \,,
\label{eq:L-E-O-tilde-prime-prime}
\end{equation}
\ie, up to $\OO(1/\MN^2)$ there are now no odd terms left and the upper and 
lower components of $\tilde{\psi}''$ are decoupled at this order.

\begin{prob}
\emph{Exercise:}  Carry out the steps that lead from the 
transformation~\eqref{eq:S-prime} to Eq.~\eqref{eq:L-E-O-tilde-prime-prime}.  
Note that it suffices to expand the exponentials up to first order.
\end{prob}

In this Lagrangian, we can now write
\begin{equation}
 \tilde{\psi}'' = \twodvec{N \\ n}
\end{equation}
and identify the upper (``large'') component $N$---a doublet in both spin and 
isospin space---with the particle and the lower (``small'') component with the 
antiparticle states.  The term $(\gamma^0-\one)\MN$ in 
Eq.~\eqref{eq:L-E-O-tilde-prime-prime} ensures that there is no explicit mass 
term for the field $N$, whereas that for $n$ comes with a factor two, 
corresponding to the Dirac mass gap between particles and antiparticles.  Let 
us now write down the Lagrangian obtained for $N$, omitting the decoupled small 
components:
\begin{equation}
 \LL_\psi = N^\dagger\left(
 \ii\partial_t 
 - g\sigma
 - \frac{1}{2\MN} \left[{-}\ii\skvec{\sigma}\cdot\vNabla
 + \ii g\skvec{\tau}\cdot\skvec{\pi}\right]
 \left[{-}\ii\skvec{\sigma}\cdot\vNabla
 - \ii g\skvec{\tau}\cdot\skvec{\pi}\right]
 \right)N + \cdots \,.
\end{equation}
To simplify this further, we use that\footnote{Note that 
Eq.~\eqref{eq:sigma-Nabla-simple} is very simple because we have not included a 
coupling of $\psi$ to the electromagnetic field.  If we had done that, the 
$\vNabla$ would be a covariant derivative, $\vec{D}=\vNabla + \ii e \vec{A}$, 
and Eq.~\eqref{eq:sigma-Nabla-simple} would generate, among other terms, the 
magnetic spin coupling $\skvec{\sigma}\cdot\vec{B}$.}
\begin{equation}
 (\skvec{\sigma}\cdot\vNabla)(\skvec{\sigma}\cdot\vNabla) = \Laplace
\label{eq:sigma-Nabla-simple}
\end{equation}
and, from the product rule,
\begin{equation}
 (\skvec{\sigma}\cdot\vNabla)(\skvec{\tau}\cdot\skvec{\pi})
 = \skvec{\sigma}\cdot(\skvec{\tau}\cdot\vNabla\skvec{\pi})
 + (\skvec{\tau}\cdot\skvec{\pi})(\skvec{\sigma}\cdot\vNabla)
 = \sigma^i\tau^a(\partial_i\pi^a) + \tau^a\pi^a \, \sigma^i\partial_i \,.
\end{equation}
In the last step we have written out all indices to clarify the meaning of the 
two dot products.  Collecting everything, we find that the 
$(\skvec{\tau}\cdot\skvec{\pi})(\skvec{\sigma}\cdot\vNabla)$ terms cancel out 
and arrive at
\begin{multline}
 \LL = N^\dagger\left(
 \ii\partial_t + \frac{\Laplace}{2\MN}\right)N
 - g\,\sigma N^\dagger N
 + N^\dagger\left(
 \frac{g}{2\MN}\skvec{\sigma}\cdot(\skvec{\tau}\cdot\vNabla\skvec{\pi})
 + \frac{g^2}{2\MN}(\skvec{\tau}\cdot\skvec{\pi})^2
 \right)N \\
 + \frac12(\partial^\mu\skvec{\pi}) \cdot (\partial_\mu\skvec{\pi})
 - \frac12\mpi^2 \skvec{\pi}^2
 + \frac12(\partial^\mu\sigma) \cdot (\partial_\mu\sigma)
 - \frac12\msigma^2 \sigma^2 + \cdots \,.
\label{eq:L-pi-N-nonrel}
\end{multline}
This includes the expected nonrelativistic kinetic term for the fermion field, 
as well as various interactions with $\sigma$ and $\skvec{\pi}$.  Note that the 
latter two particles are still relativistic and unchanged by the 
Foldy-Wouthuysen transformation, so that we could simply reinstate their 
kinetic terms as in Eq.~\eqref{eq:L-pi-N-PS}.

\subsubsection{Integrating out exchange particles: part II}
\label{sec:EFT-IntOut-2}

With Eq.~\eqref{eq:L-pi-N-PS} we are now also in a convention situation to 
illustrate how we end up with only contact interactions between the 
nonrelativistic fermions if we integrate out the $\sigma$ and $\skvec{\pi}$ 
fields.  Their equations of motion are
\begin{equation}
 \left(\dAlem + \msigma^2\right)\sigma = g\,N^\dagger N
\label{eq:sigma-N-EOM-NR}
\end{equation}
and
\begin{equation}
 \left(\dAlem + \mpi^2\right)\pi^\lambda
 = {-}\frac{g}{2\MN} 
 \vNabla\cdot \left[N^\dagger\skvec{\sigma}\tau^\lambda N\right]
 - \frac{g^2}{\MN}
 N^\dagger (\skvec{\tau}\cdot\skvec{\pi}) \tau^\lambda N \,.
\label{eq:pi-N-EOM-NR}
\end{equation}
The $\sigma$ part can be handled exactly as in Sec.~\ref{sec:EFT-IntOut-1}, 
giving a leading four-nucleon contact interaction $\sim g^2/\msigma^2$ plus a 
tower of operators with increasing number of derivatives.  The $\skvec{\pi}$ 
part is more interesting, but also more complicated due to the derivative in 
Eq.~\eqref{eq:pi-N-EOM-NR}.  We thus keep the following discussion rather 
qualitative and leave it as an exercise to work out the details.

In that spirit, we consider only the first term in Eq.~\eqref{eq:pi-N-EOM-NR},
corresponding to a one-$\skvec{\pi}$-exchange operator when substituted back 
into the Lagrangian.  With the propagator $D_\pi(x-y)$ defined in complete 
analogy to Eq.~\eqref{eq:D-chi-KG}, we can write
\begin{equation}
 \pi^\lambda(x) = {-}\frac{\ii g}{2\MN} \int\dd^4y \, D_\pi(x-y)
 \, \partial_j^y \left[N^\dagger(y)\,{\sigma^j}\tau^\lambda N(y)\right]
 + \cdots \,,
\end{equation}
and thus get
\begin{equation}
 \LL_\text{int}
 \sim \left[N^\dagger(x)\,\sigma^i\tau^\lambda N(x)\right]
 \partial_i^x \int\dd^4y \, D_\pi(x-y)
 \, \partial_j^y \left[N^\dagger(y)\,\sigma^j\tau^\lambda N(y)\right]
 + \cdots \,,
\end{equation}
where for the time being we omit the prefactor ${g^2}/{(4\MN^2)}$.  We 
integrate by parts to have $\partial_j^y$ act on $D_\pi(x-y)$.  The 
$\partial_i^x$ does this already, so, with all indices written out for clarity:
\begin{multline}
 \LL_\text{int} \sim \int\dd^4y \,
 N^\dagger_{\alpha a}(x)\, \idxx{\sigma^i}{\alpha}{\beta}
 \idxx{\tau^\lambda}{a}{b} \, N^{\beta b}(x)
 \left[\partial_x^i \partial_y^j D_\pi(x-y) \right]
 N^\dagger_{\gamma c}(y) \idxx{\sigma^j}{\gamma}{\delta}
 \idxx{\tau^\lambda}{c}{d} \, N^{\delta d}(y)
 + \cdots \,.
\label{eq:L-pi-N-Dpi}
\end{multline}
From the definition of the propagator we find that
\begin{equation}
 \partial^x_i \partial^y_j D_\pi(x-y)
 = \int\frac{\dd^4p}{(2\pi)^4} (\ii p_i)({-}\ii p_j) \eex^{{-}\ii p(x-y)}
 \frac{\ii}{p^2-\mpi^2 + \ii\eps} 
 = {-}\partial^x_i \partial^x_j D_\pi(x-y) \,,
\label{eq:D-pi-partial}
\end{equation}
so the partial derivatives can be written fully symmetric in $i$ and $j$.  The 
various fermion field operators can be rearranged with the help of
\begin{multline}
 N^{\beta b}(x) N^\dagger_{\gamma c}(y)
 = {-}\frac{1}{4}\Big[
 (N^\dagger(y) N(x))\,\delta^\beta_\gamma \delta^b_c
 + (N^\dagger(y) \tau^\kappa N(x))
 \,\delta^\beta_\gamma \idxx{\tau^\kappa}{b}{c} \\
 + (N^\dagger(y) \sigma^k N(x))\,\idxx{\sigma^k}{\beta}{\gamma} \delta^b_c
 + (N^\dagger(y)\,\tau^\kappa\sigma^k N(x)) \,
 \idxx{\sigma^k}{\beta}{\gamma}\idxx{\tau^\kappa}{b}{c} \Big] \,.
\label{eq:NN-Fierz}
\end{multline}
Using also
\begin{subalign}
 \sigma^i\sigma^j &= \delta^{ij}\one + \ii\leviciv^{ijk}\sigma^k \,, \\
 \sigma^i\sigma^k\sigma^j &= \delta^{kj}\sigma^i + \delta^{ki}\sigma^j
 - \delta^{ij}\sigma^k + \ii\leviciv^{ikj}\one \,, \\
  \sigma^i\sigma^j\sigma^i &= {-}\sigma^j \,,
\end{subalign}
we get four terms from Eq.~\eqref{eq:L-pi-N-Dpi} decomposed into contributions 
symmetric and antisymmetric in $i$ and $j$, with the latter all vanishing upon 
contraction with $\partial^x_i \partial^x_j$.  The simplest symmetric term 
comes with a $\delta^{ij}$, yielding $\Laplace D_\pi(x-y)$.  To see what this 
generates, we Taylor-expand the fermion fields that depend on $y$ about $x$, 
\eg, $N(y) = N(x) + (y-x)^\mu \partial_\mu N(x) + \cdots$.  This gives as the 
leading piece a combination of four fermion operators all evaluated at $x$, 
times
\begin{equation}
 \int\dd^4y \int\frac{\dd^4p}{(2\pi)^4} \eex^{{-}\ii p\cdot(x-y)}
 \frac{\ii{\mathbf{p}}^2}{p^2-\mpi^2 + \ii\eps} \\
 = \int\dd^4y \int\frac{\dd^3p}{(3\pi)^4} \int\frac{\dd p_0}{2\pi} \,
 \eex^{{-}\ii p(x-y)}
 \frac{\ii{\mathbf{p}}^2}{p_0^2 - {\mathbf{p}}^2 - \mpi^2 + \ii\eps} \,.
\end{equation}
The integral over $p_0$ can be solved via contour integration.  Defining
$\omega_{\mathbf{p}} = \sqrt{{\mathbf{p}}^2+\mpi^2}$, we get
\begin{equation}
 \int\frac{\dd p_0}{2\pi} \, \eex^{{-}\ii p_0(x_0-y_0)}
 \frac{\ii}{p_0^2 - {\mathbf{p}}^2 - \mpi^2 + \ii\eps}
 = \frac{\eex^{-\ii\omega'_{\mathbf{p}}\abs{x_0-y_0}}}{2\omega'_{\mathbf{p}}}
 \mathtext{with}
 \omega'_{\mathbf{p}} = \omega_{\mathbf{p}} - \frac{\ii\eps}{2\omega_{\mathbf{p}}}
 \equiv \omega_{\mathbf{p}} - \ii\eps' \,.
\end{equation}
It is important here to keep track of the small imaginary part, as it allows us 
to write
\begin{equation}
 \int_{{-}\infty}^\infty \dd y_0 \,
 \frac{\eex^{-\ii\omega_{\mathbf{p}}'\abs{x_0-y_0}}}{2\omega'_{\mathbf{p}}}
 = \int_{0}^\infty \dd y_0
 \frac{\eex^{-\ii\omega'_{\mathbf{p}}\abs{y_0}}}{\omega'_{\mathbf{p}}}
 = {-}\ii / {(\omega'_{\mathbf{p}})^2} \,.
\end{equation}
Collecting the results up to this point, we arrive 
at
\begin{equation}
 \int\dd^4y \int\frac{\dd^4p}{(2\pi)^4}
 \, \eex^{{-}\ii p(x-y)}
 \dfrac{\ii{\mathbf{p}}^2}{\strut p^2 - \mpi^2 + \ii\eps}
 = {-}\!\int\dd^3y \int\frac{\dd^3p}{(2\pi)^3}
 \, \eex^{\ii {\mathbf{p}}\cdot(\vx-\vy)}
 \frac{{\mathbf{p}}^2}{\strut{\mathbf{p}}^2+\mpi^2} \,.
\end{equation}
We finally obtain the desired contact interaction by expanding
\begin{equation}
 \frac{{\mathbf{p}}^2}{\strut{\mathbf{p}}^2+\mpi^2}
 = \frac{{\mathbf{p}}^2}{\mpi^2} \left(1 - \frac{{\mathbf{p}}^2}{\mpi^2} + \cdots\right) \,,
\end{equation}
with a leading term $\sim{\mathbf{p}}^2$, generating a contact interaction $\sim 
(N^\dagger N)\Laplace(N^\dagger N)$.  This is of course not surprising: after 
all, the original interaction term in Eq.~\eqref{eq:L-pi-N-nonrel} generating 
the contact operator had a single derivative $\vNabla$.  Considering other 
terms coming from Eq.~\eqref{eq:NN-Fierz}, one can also find operators like 
$(N^\dagger \skvec{\sigma}\cdot\vNabla N)(N^\dagger \skvec{\sigma}\cdot\vNabla 
N)$, and it is a useful exercise to work this out in detail.  But already from 
our qualitative discussion here we can infer that the resulting effective 
theory is an expansion in ${\mathbf{p}}^2/\mpi^2$, \ie, its range of validity is 
determined by three-momenta---rather than the energies---being small 
compared to $\mpi$.\footnote{This is assuming $\mpi<\msigma$.}

\subsubsection{The Schr\"odinger field}

We conclude this section by looking at the non-relativistic field theory from
a more general perspective, establishing its close connection to the ``second
quantized'' approach to (many-body) quantum mechanics that is used 
in several later chapters of this volume.

Recall from the beginning of this section that the 
Lagrangian~\eqref{eq:L-phi-free-nonrel} for the free Schr\"odinger field 
$\phi$ is
\begin{equation}
 \LL_{\phi,\text{free}} = \phi^\dagger
 \left(\ii\partial_t + \frac{\Laplace}{2m}\right)\phi \,.
\label{eq:L-phi-free-nonrel-hat}
\end{equation}
This trivially gives the equation of motion
\begin{equation}
 \left(\ii\partial_t + \frac{\Laplace}{2m}\right)\phi = 0 \,,
\label{eq:phi-nonrel-free-EOM}
\end{equation}
which is formally the same as the free Schr\"odinger equation.  However, recall 
that $\phi$ here is a field \emph{operator}, \ie, $\phi(x)$ creates a particle 
at $x=(t,\vx)$ from the vacuum, so to really get an ordinary Schr\"odinger 
equation, we have to act with both sides of Eq.~\eqref{eq:phi-nonrel-free-EOM} 
on $\ket{0}$, and define the quantum-mechanical one-body state
\begin{equation}
 \ket{\phi(t,\vx)} = \phi(t,\vx)\ket{0} \,.
\end{equation}
If we add to Eq.~\eqref{eq:L-phi-free-nonrel-hat} a term 
$V(x)\phi^\dagger(x)\phi(x)$, we obtain the Schr\"odinger equation for a 
particle in a potential $V(x)$.  Exactly as for a relativistic field we can 
define the propagator
\begin{equation}
 D_\phi(x-y)
 = \int\frac{\dd^4q}{(2\pi)^4} \, \eex^{{-}\ii p(x-y)}
 \frac{\ii}{p_0-\frac{{\mathbf{p}}^2}{2m} + \ii\eps} \,,
\label{eq:D-phi}
\end{equation}
satisfying
\begin{equation}
 \left(\ii\partial_t + \frac{\Laplace}{2m}\right)D_\phi(x-y)
 = {-}\ii \delta^{(4)}(x-y) \,.
\end{equation}
Up to a conventional factor $\ii$, this is precisely the (retarded) Green's 
function\footnote{Note that in the nonrelativistic case there is no ``Feynman 
propagator.''  Particles and particles are decoupled, and the denominator in 
Eq.~\eqref{eq:D-phi} only has a single pole at $p_0=\vec{{\mathbf{p}}^2}/(2m)-\ii\eps$.  
Flipping the sign of the $\ii\eps$ term gives the advanced Green's function.} 
familiar, for example, from non-relativistic scattering theory (then typically 
denoted $G_0$).  This will appear again when the Lippmann--Schwinger equation 
is derived using the field-theory language in Sec.~\ref{sec:EFT-Bosons}.

While it is nice and reassuring that we can go back to simple quantum mechanics 
from the one-body Schr\"odinger Lagrangian discussed so far, this feature is 
not very relevant in practice.  We can, however, straightforwardly generalize 
it to the many-body case.  To that end, consider a Lagrangian that includes a 
two-body interaction, written in terms of a general non-local 
potential:\footnote{A static (time-independent) potential, as it is more common 
in quantum mechanics, would be a function only of $\vx$ and $\vy$, and all 
fields in the interaction term would be evaluated at the same time $t$.}
\begin{equation}
 \LL_{\phi,\text{2-body}}(x) = \phi^\dagger(x)
 \left(\ii\partial_t + \frac{\Laplace}{2m}\right)\phi(x)
 + \int\dd^4 y \, \phi^\dagger(x) \phi(x)
 \, V(x,y) \, \phi^\dagger(y) \phi(y) \,.
\label{eq:L-phi-V2-nonrel-hat}
\end{equation}
Note that this has exactly the structure that we found when we integrated out 
particles in the preceding sections, before expanding the propagators to get 
simple contact interactions.  Such a Lagrangian (possibly including also 
higher-body forces) is a convenient starting point for example for many-body 
perturbation theory used to study quantum systems at finite density.

Coming back to effective field theories, we stress that these are \emph{not} 
defined by putting a given potential into a Lagrangian; in doing that, 
one merely gets a model written in a convenient way.  The EFT instead makes no 
assumptions on the interaction (besides symmetry constraints).  It is thus much 
more general and not a model, but to be predictive it requires a number of 
\apriori unknown parameters to be fixed and its various terms to be ordered 
systematically.  It is this that we turn to next.

\subsection{Symmetries and power counting}

So far, we have discussed how to obtain effective low-energy Lagrangians by 
integrating out "heavy" degrees of freedom, leaving only those that we want  
to describe at low energies or rather, as we showed explicitly with the 
pseudoscalar pion-nucleon model, low momenta.  We found the contact 
interactions generated this way to come with the integrated-out particle's mass 
in the denominator, and with an increasing number of derivatives as we keep 
more and more terms from the expansion.  These derivatives will turn into 
powers of momentum, which is a small scale for external states.  We furthermore 
showed how a nonrelativistic reduction generates a chain of operators with an 
increasing power of the particle's mass in the denominator, thus also giving a 
hierarchy of terms that eventually restore the original theory's relativistic 
structure with coupling between particles and antiparticles.

From these procedures it is clear that the terms in the effective Lagrangian 
should be ordered in a natural way, with the most important ones being those 
with the least number of large mass scales in the denominator and the least 
number of derivatives in the numerator.  It is also clear that they are 
restricted in their structure.  For example, if we start with a 
Lorentz-invariant relativistic theory, after the nonrelativistic reduction we 
will only get terms that are invariant under ``small'' Lorentz boosts.  More 
precisely, the nonrelativistic operators should be invariant under Galilean 
transformations (assuming the original theory had rotational invariance, this 
simply gets inherited by the effective one), and the form of so-called 
``relativistic corrections'' is determined by the expansion of the 
dispersion relation for positive-energy solutions:
\begin{equation}
 E = \sqrt{m^2 + p^2}
 = m + \frac{p^2}{2m} - \frac{p^4}{8m^3} + \cdots \,.
\end{equation}

We now turn to discussing the bottom-up approach guided by these principles.  
To that end, consider the effective Lagrangian for a nonrelativistic bosonic 
field with contact interactions:
\begin{multline}
 \LL = \phi^\dagger \left(\ii\partial_t + \frac{\Laplace}{2m}\right)\phi
 + \phi^\dagger \frac{\vNabla^4}{8m^3} \phi + \cdots \\
 + g_{2}^{(0)} (\phi^\dagger \phi)^2
 + g_{2}^{(2s/p)} \left(
  (\phi^\dagger\vNablaLR\phi)^2
   - (\phi^\dagger\phi)(\phi^\dagger(\vNablaLR)^2\phi)
   \mp 2 (\phi^\dagger\phi)\vNabla^2(\phi^\dagger\phi)
  \right) + \cdots \\
 + g_{3}^{(0)} (\phi^\dagger \phi)^3 + \cdots \,.
\label{eq:L-phi-generic}
\end{multline}
Here we have used the definition
\begin{equation}
 f \vNablaLR g = f(\vNabla)g - (\vNabla f)g
\end{equation}
and conveniently separated the two-body terms with two derivatives into 
those which contribute to S-wave ($\sim g_{2}^{(2s)}$) and P-wave ($\sim 
g_{2}^{(2p)}$) interactions, respectively.  One can of course choose different 
linear combinations, but a separation by partial waves is typically a good 
choice for systems with rotational invariance.  It is a useful exercise to work 
out how the structure for the derivative interactions gives the desired result, 
working in momentum space and considering contractions with external in and out 
states that have center-of-mass momenta $\pm\vk^2$ and $\pm{\mathbf{p}}/2$, respectively.
The structure of the individual terms is determined by the requirement of 
Galilean invariance,\footnote{See for example Ref.~\cite{Hagen:2014}, 
Sec.~2.1.1 for a rigorous discussion of the required transformation 
properties.}, and the EFT paradigm tells us to write down all possible terms 
with a given number of derivatives (with odd numbers excluded by parity 
invariance).

\subsubsection{The breakdown scale}

As mentioned in the introduction, the most important requirement to construct 
an EFT is the identification of---at least two, but possibly more---separated 
scales, ratios of which are used to extract a small expansion parameter.  The 
better the scale separation, the smaller this parameter becomes, and 
consequently the better the more precise (and, provided all contributions have 
been identified correctly, accurate) the theory becomes at any given order in 
the expansion.  In the simplest case, there is one low scale $Q$ 
associated with the typical momentum of the physical system that we want to 
describe, and a single large scale $\Mhi$, the ``breakdown scale'' associated 
with the physics that our EFT does not take into account---in other words: 
resolve---explicitly.  This is exactly the situation that we constructed when we 
integrated out exchange particles from a given theory in 
Secs.~\ref{sec:EFT-IntOut-1} and~\ref{sec:EFT-IntOut-2}.  By construction, the 
EFT is not appropriate to describe processes with momenta of the order of or 
large than the breakdown scale.  To emphasize this meaning, it is sometimes also 
denoted by the letter $\Lambda$ (with or possibly without some qualifying 
subscript).\footnote{We alert the reader that in the literature this is 
sometimes referred to as the ``cutoff of the EFT.''  We do not use that language 
to avoid confusion with an (arbitrary) momentum cutoff introduced to regularize 
divergent loop integrals (discussed .}

As already mentioned, integrating out degrees of freedom from a given more 
fundamental theory will naturally give a breakdown scale set by that particle's 
mass.  But it can also be something more general.  For example, although in the 
situations discussed here so far the particles we were ultimately interested in 
were already present as degrees of freedom in the original theory, such a 
scenario is merely a special case.  The first step in writing down an effective 
field theory is to identify what the appropriate---literally: 
effective---degrees of freedom are for the processes one wants to describe, and 
they can be different from those of the fundamental theory.  This is exactly 
the case in nuclear physics: while the degrees of freedom in quantum 
chromodynamics (QCD) are quarks and gluons, describing the binding of nuclei 
with these is, although possible with state-of-the-art lattice QCD 
calculations, largely inefficient to say the least.  It is much more economical 
to work with nucleons directly as degrees of freedom, as done in most chapters 
of this volume, because a detailed knowledge of the internal structure of 
protons and neutrons is not necessary to describe their binding into nuclei; it 
is only resolved at much higher energies, for example in deep inelastic 
scattering.  The reason for this is color confinement: the low-energy degrees 
of freedom of QCD are not quarks and gluons, but color-neutral hadrons.  Chiral 
effective field theory, which we will come back to in 
Sec.~\ref{sec:EFT-Chiral}, is designed to work at momenta of the order of pion 
mass, breaking down at the scale of chiral-symmetry breaking (estimated to be 
roughly a \GeV, but possibly lower).

Other examples are halo EFT, constructed to describe nuclear systems that have 
the structure of a few nucleons weakly bound to a tight core, which can then 
effective be treated as a structureless particle.  Clearly, such a theory will 
break down at momenta large enough to probe the core's internal structure.  
Similarly, one can construct an effective theory for systems of ultracold 
atomic gases, the constituents of which can be treated as pointlike degrees of 
freedom without using QED to describe their individual structure, and much less 
QCD to describe their atomic nuclei.

Whatever the breakdown scale is, once identified it can be used to 
systematically order terms in the effective Lagrangian by powers of 
$Q/\Mhi \ll 1$, and we now turn to discussing how this ordering can be set 
up.

\subsubsection{Na\"ive dimensional analysis}

In our units with $\hbar = c = 1$, the action
\begin{equation}
 S = \int \dd^4 x\, \LL(x)
\end{equation}
has to be a dimensionless quantity.  This, in turn, fixes the dimensions for 
the individual building blocks in the Lagrangian.  In a relativistic theory,  
mass and energy are equivalent and one would simply express everything in terms 
of a generic mass dimension.  For our nonrelativistic framework, on the other 
hand, energies are \emph{kinetic} energies because the time dependence 
associated with the rest mass has been absorbed into the field (\cf 
Secs.~\ref{sec:EFT-NonRelBos} and~\ref{sec:EFT-NonRelFerm}).  This implies 
that energy and mass scales---as well as time and space---should be counted 
separately.\footnote{This separation would be quite clear if we had not set 
$c=1$, which would in fact be more appropriate for a nonrelativistic system.  
The reason we still do it that it allows us to still energies and momenta in the 
same units, \eg, in $\MeV$, following the standard convenient in nuclear 
physics.}  In fact, it is more natural to consider powers of momentum.  To 
understand what this means, let us start with the kinetic term in 
Eq.~\eqref{eq:L-phi-generic}: $[\vNabla^2/(2m)] = 
\text{momentum}^2/\text{mass}$.  The time derivative has to scale in the same 
way, implying that for time itself we have $[t] = 
\text{mass}/\text{momentum}^2$, whereas $[x] = \text{momentum}^{-1}$.  
Consequently, the integration measure scales like $[\dd^4 x = \dd t\,\dd^3x] = 
\text{mass}/\text{momentum}^5$ (to compare, in the relativistic theory one 
would simply count $[\dd^4 x] = \text{mass}^{-4} = \text{energy}^{-4}$).

Since the dimension of $\LL$ has to cancel that of the measure to give a 
dimensionless action,we can now infer that our field has to satisfy $[\phi] = 
\text{momentum}^{3/2}$, \ie, even though it is a scalar field it scales with a 
fractional dimension (recall that in the relativistic case a scalar would have 
dimension $\text{energy}^1$).  Knowing the scaling of the field and the 
measure, we can now proceed and deduce that of the various coupling constants.

The basic idea is very simple: each term (operator) in the 
Lagrangian~\eqref{eq:L-phi-generic} has $2n$ fields and $2m$ derivatives, 
giving it a total dimension of $\text{momentum}^{3n+2m}$.  For example, the 
$(\phi^\dagger \phi)^2$ term with $2n=4$ and $m=0$ has dimension 
$\text{momentum}^6$.  Hence, to get the correct overall dimension 
$\text{momentum}^5/\text{mass}$ for $\LL$, the coupling constant $g_{2}^{(0)}$ 
has to be $\sim 1/(\text{momentum}\times\text{mass})$.  Since it is supposed to 
describe unresolved short-distance details, the momentum scale in the 
denominator is should be the breakdown scale, whereas the mass scale, which 
as we mentioned is a feature of the nonrelativistic framework and common to 
all operators, is simply associated with $m_\phi$.  Of course, counting a 
single operator does not tell us much: it is the relative order of terms that 
matters, so we proceed to the $g_2^{(2)}$ interactions.  These all come with 
two derivatives, which are associated with the external (small) momentum scale 
$Q$. Hence, we have $2n=4$ and $2m=2$, and we need to compensate the two 
additional powers of momentum in the numerator with two more powers of 
$\Mhi$ in the denominator, finding that the $g_2^{(2)}$ interactions are 
down compared to the $g_2^{(0)}$ term by a factor $(Q/\Mhi)^2$.  This is 
exactly in line with our picture of the contact terms gradually building up the 
an unresolved particle exchange through a derivative expansion.  For 
higher-body interactions, it is the larger number of fields that gives a 
suppression by inverse powers of $\Mhi$ compared to operators with fewer 
fields.

This kind of analysis can be much improved if something is known about which 
unresolved physics is supposed to be represented by which operator, and it is 
generally more complex if the theory involves different fields.  For example, in 
the EFT for halo nuclei there are contact interactions associated with 
unresolved pion exchange, as well as those systematically accounting for the 
internal structure of the core field.  Instead of merely putting generic powers 
of $\Mhi$ in every denominator, it can be necessary to keep track of several 
high scales separately to figure out the ordering of terms.  Also, it is 
possible that the external momentum is not the only relevant low-momentum scale 
in the problem.

This rather abstract discussion will become clearer when we finally discuss 
concrete EFTs in the following sections~\ref{sec:EFT-Bosons} 
and~\ref{sec:EFT-Nucleons}.  In that context, we will use the scaling of 
the various terms in the Lagrangian to power-count diagrams as a 
whole, \ie to estimate the size of individual contributions composed of 
vertices and loops to a given physical amplitude of interest.  We will then 
also discuss how the actual so-called scaling dimension of a field in the 
Lagrangian can turn out to deviate from what we estimated here based purely on 
dimensional grounds.

\subsubsection{Fine tuning}
\label{sec:EFT-FineTuning}

In connection with the previous comment is another point worth stressing already 
here: na\"ive dimensional analysis resides at the beginning of EFT wisdom, not 
at the end, and in quite a few cases it turns out to be exactly what the name 
says: na\"ive.  In other words, the actual scaling of a coupling constant can be 
quite different from what one would infer by counting dimensions, a scenario 
that is commonly referred to as ``fine tuning.''  To understand why that is 
consider, for example, our bosonic toy model from Sec.~\ref{sec:EFT-IntOut-1}, 
but now assume that there already is a four-$\phi$ contact interaction present 
prior to integrating out the $\chi$ field:
\begin{equation}
 \LL = {-}\phi^\dagger\left(\dAlem + m_\phi^2\right)\phi
 - \chi^\dagger\left(\dAlem + m_\chi^2\right)\chi
 + g\left(\phi^\dagger\phi^\dagger\chi + \hc\right)
 + h (\phi^\dagger\phi)^2 \,.
\label{eq:L-phi-chi-rel-contact}
\end{equation}
This could, for example, come from unknown (or integrated out) short-distance 
physics at a yet higher scale.  When we now integrate out the $\chi$, the 
generated non-derivative contact term will combine with the existing one, 
giving a single operator in the effective low-energy Lagrangian (recall that on 
dimensional grounds $h$ has to have dimensions of inverse mass squared):
\begin{equation}
 \LL = {-}\phi^\dagger\left(\dAlem + m_\phi^2\right)\phi
 + \left(h-\frac{g^2}{m_\chi^2}\right) (\phi^\dagger\phi)^2 + \cdots \,.
\label{eq:L-phi-chi-rel-contact-FT}
\end{equation}
Now suppose we had started in the bottom-up approach and simply written down 
the four-$\phi$ contact operator with some coefficient $c$ to be determined.  
According to NDA, we would assume that its scale is set by two powers of the 
breakdown scale in the denominator, and assuming we actually know about the 
more fundamental theory, we might have estimated that breakdown scale to be of 
the order $m_\chi$.  From Eq.~\eqref{eq:L-phi-chi-rel-contact-FT} we see 
that depending on what values $g$ and $h$ take in the underlying theory, the 
actual size of $c$ might deviate strongly from the na\"ive expectation, and it 
could even be set by a low-energy scale of the effective theory.  But for this 
to happen, there would have to be a delicate cancellation between $h$ and 
$g^2/m_\chi^2$, which is typically deemed unlikely given the \apriori vast 
range of possible values these parameters could take; thus the term ``fine 
tuning.''  The fact that coupling constants are in fact not simple fixed 
numbers but get renormalized by loop effects (\ie, depend on a regularization 
scale with a behavior determined by the renormalization group) justifies this 
language even more.

\subsubsection{Loops and renormalization}
\label{sec:Loops}

It is indeed high time we talk about loops.  Our considerations in this section 
so far have been limited to tree level, which is always only a first 
approximation in a quantum field theory.  In a perturbative theory, loop 
contributions from virtual intermediate states are added to improve the 
accuracy of the result.  To treat a nonperturbative system such as a bound 
nucleus, on the other hand, they are absolutely crucial: recall that any finite 
sum of diagrams in perturbation can never produce a bound state (for example, 
think about poles in the S-matrix, which cannot be generated through a finite 
sum of terms).  In the field-theory language, this means that an infinite 
number of diagrams with increasing number of loops has to be summed to get the 
amplitude with the desired physical properties.

This situation is in fact familiar already from the Schr\"odinger equation 
written in the form
\begin{equation}
 \ket{\psi} = \hat{G}_0(E) \hat{V} \ket{\psi} \mathtext{,}
 \hat{G}_0 = \hat{G}_0(E) = (E - \hat{H}_0)^{-1} \,,
\end{equation}
which can be iterated to get $\ket{\psi} = \hat{G}_0\hat{V}
\hat{G}_0\hat{V} \ket{\psi} = \cdots$.  When these operators are written 
out in momentum space, each propagator $\hat{G}_0$ corresponds to a loop.
More closely related to the amplitude written down in a nonrelativistic field 
theory, this exercise can be repeated with the Lippmann-Schwinger equation for 
the T-matrix and its formal solution, the infinite Born series.  Exactly this 
will be recovered in Sec.~\ref{sec:EFT-Bosons}.

Of course, even in a nonperturbative theory we do not expect that \emph{all} 
loop diagrams should be summed up to infinity.  Generally, we want the power 
counting to tell us how to estimate the contribution from a given diagram, 
including loop diagrams.  To do that, we need to know not only what which 
factors we pick up from vertices, but also need an estimate for the integration 
measure $\dd^4q = \dd q_0\,\dd^3 q$.  Any loop diagram contributing to an 
amplitude with external momenta of the order $Q$ will have this scale running 
through it a whole.  It is thus natural to count the contribution from the 
three-momentum as $\dd^3 q \sim Q^3$ and, recalling that in the nonrelativistic 
theory $q_0$ is a kinetic energy, $\dd q_0 \sim Q^2/m$.  For each Schr\"odinger 
propagator we get, conversely, a factor $m/Q^2$, as can be seen from 
Eq.~\eqref{eq:D-phi}.  These simple rules combined with those for the vertices 
give an estimate for any diagram in the theory determined by 
Eq.~\eqref{eq:L-phi-generic}.

What this discussion does not cover is the fact that loops in a quantum field 
theory can be---and mostly are---divergent.  Compared to the loops one gets 
from integrating the Schr\"odinger or Lippmann-Schwinger equation in quantum 
mechanics with a potential $\hat{V}$, which are typically all finite, this is 
different in the EFT simply because our delta-function (contact) interactions 
are too singular to make direct sense beyond tree level.  Of course, this is no 
different than in any other quantum field theory, and it just means that 
divergent loops have to be regularized (for example, by imposing a momentum 
cutoff or with dimensional regularization), and then suitable renormalization 
conditions have to be imposed to fix the various coupling constants in the 
effective Lagrangian.  These then become functions of the renormalization scale, 
with a behavior governed by the renormalization group (RG).  In 
Sec.~\ref{sec:EFT-Bosons} this will be discussed in detail for a bosonic EFT 
that describes, for example, ultracold atomic systems.

\paragraph{The cutoff}

What the regularization of loop integrals does is most transparent with a 
momentum cutoff.  We denote this by $\Lambda$ and stress again that it has to 
be distinguished from the EFT breakdown scale $\Mhi$.  The latter determines 
the scale beyond which we know our EFT not to be valid.  In other words, 
short-range dynamics corresponding to momenta larger than $\Mhi$ is, in 
general, not correctly described by the EFT.  Yet from loop integrals we get 
contributions from states up to the UV cutoff $\Lambda$.  Renormalization means 
to adjust the coupling constants in such a way that they compensate the wrong 
high-momentum loop contributions in such a way that the physics the EFT is 
supposed to describe comes out correctly.  For momenta up to $\Mhi$, we 
trust the EFT, so it makes sense to keep such states in loops.  Hence, one 
should typically choose $\Lambda > \Mhi$.  Choosing it lower than the 
breakdown scale is possible, but this can induce corrections of the order 
$Q/\Lambda > Q/\Mhi$, which is not desirable for the power counting.  In the 
renormalized EFT, any cutoff in the interval $[\Mhi,\infty)$ is thus an 
equally good choice---it does not have to be ``taken to infinity.''  Instead, 
that phrase should be understood to mean adjusting the couplings at any given 
finite cutoff.  If this procedure is carried out numerically, it can be 
desirable to keep the cutoff small, but one has to make sure that in principle 
in \emph{can} be varied arbitrarily.

\subsection{Matching}

The determination of the couplings (``low-energy constants'') in the effective 
Lagrangian is done by expressing a given physical quantity (\eg, a scattering 
amplitude or related ) in terms of the couplings and then adjusting them to 
reproduce a known result.  This can be done using experimental input or, when 
working top-down, by calculating the same amplitude in the more fundamental 
theory.  Generally, this procedure is referred to as ``matching.''  At tree 
level, this is again exactly what we did by integrating out particles and found 
the coefficients of the generated contact interactions in terms of the original 
coupling and mass denominators.  Once loop diagrams are involved, the process 
becomes somewhat more complicated because (a) one has to make sure to use 
compatible regularization schemes and renormalization scales and (b) loop 
diagrams with lower-order vertices typically mix with higher-order tree-level 
diagrams.  The latter is a general feature of combined loop and derivative 
expansions and is thus also important when matching to experimental input.  
While these comments may sound a bit cryptic here, they will become much clearer 
in the next section when we finally work with a concrete EFT.

\section{Effective field theory for strongly interacting bosons}
\label{sec:EFT-Bosons}

We will now use the insights from the previous sections to
construct a local effective field theory for 
identical, spinless bosons with short-range S-wave 
interactions.\footnote{See Ref.~\cite{Braaten:2004rn} for a similar 
discussion with a focus on applications in ultracold atoms.}
For the treatment of higher partial wave interactions the 
reader is referred to the 
literature~\cite{Beane:2000fx,Bertulani:2002sz,Bedaque:2003wa}. 
The most general effective Lagrangian consistent with 
Galilei invariance can be written as
\begin{equation}
 {\mathcal L} = \phi^\dagger \left(\ii\partial_t + \frac{\nabla^2}{2m} 
 \right)\phi - \frac{C_0}{4} \left(\phi^\dagger \phi
 \right)^2 - \frac{C_2}{4} \left(\nabla (\phi^\dagger \phi)\right)^2
 + \frac{D_0}{36}  \left(\phi^\dagger \phi \right)^3+\cdots \,.
\label{L-2body}
\end{equation}
where $m$ is the mass of the particles and
the ellipses denote higher-derivative and/or higher-body
interactions. The leading two- and three-body interactions are 
explicitly written out. The scaling of the coefficients $C_0$, $C_2$,
$D_0$, $\ldots$ depends on the scales of the considered
system. Two explicit examples, corresponding natural and unnaturally
large scattering length, are discussed below.

\subsection{EFT for short-range interactions}

We start by considering natural system where all interactions 
are characterized by only one mass scale $\Mhi$ that we identify with the 
formal breakdown scale of the EFT introduced in the previous section.  We will 
see below that this is indeed justified.  Since the nonrelativistic boson fields 
have dimension 3/2, the coupling constants must scale
as
\begin{equation}
 C_0 \sim \frac{1}{m\Mhi}\,,\quad C_2 \sim \frac{1}{m\Mhi^3}\,,
 \quad\text{and}\quad
 D_0 \sim \frac{1}{m\Mhi^4}\,,
\label{eq:cscale}
\end{equation}
such that higher dimension operators are strongly suppressed for small 
momenta $k \ll \Mhi$.\footnote{Note that coupling constants scale 
with the particle mass as 
$1/m$ in nonrelativistic theories. This can be seen by rescaling all 
energies as $q_0 \to \tilde{q_0}/m$ and all time coordinates as 
$t\to \tilde{t}m$, so that dimensionful quantities are measured in 
units of momentum. Demanding that the action is independent of $m$,
it follows that the coupling constants must scale as $1/m$.}
We first focus on the two-body system and
calculate the contribution of the interaction terms in Eq.~(\ref{L-2body}) 
to the scattering amplitude of two particles in perturbation theory. 
After renormalization, the result reproduces
the low-energy expansion of the 
scattering amplitude for particles with relative momentum $k$ and 
total energy $E=k^2/m$:
\begin{equation}
 T_2(E)=\frac{8\pi}{m}\frac{1}{k\cot\delta_0(k)-\ii k}=
 {-}\frac{8\pi a}{m}\left(1-\ii ak+ (ar_e/2-a^2)k^2 +{\mathcal O}(k^3)\right)\,,
\label{eq-ere}
\end{equation}
where the effective range expansion for short-range interactions
$k\cot\delta_0(k)={-}1/a+r_e k^2/2 +{\mathcal O}(k^4)$ has been used.

Since all coefficients of the effective Lagrangian are natural (scaling with 
inverse powers of $\Mhi$), it is sufficient to count the powers of small momenta 
$Q$ in scattering amplitudes to determine the scaling of the amplitudes with 
$\Mlo$. The correct dimensions are made up with appropriate factors of $\Mhi$ 
contained in the coupling constants (\cf~Eq.~(\ref{eq:cscale})).
For a general two-body amplitude with $L$ loops and $V_{2i}$ interaction
vertices with $2i$ derivatives, we thus have $T_2 \sim Q^\nu$ where the
power $\nu$ is given by
\begin{equation}
  \nu=3L+2+\sum_i (2i-2) V_{2i} \geq 0\,.
\label{nu-pert}
\end{equation}
Here we have used that loop integrations contribute a factor $k^5$
and propagators a factor $k^{-2}$ in nonrelativistic theories.
The values of the coupling constants $C_0$ and $C_2$ can be determined by 
matching to Eq.~(\ref{eq-ere}). In the lowest two orders only $C_0$ 
contributes.

\begin{prob}
{\emph Exercise:}  Derive Eq.~(\ref{nu-pert}) using the topological
identity for Feynman diagrams:
\begin{equation}
 L=I-V+1\,,
\end{equation}
with $L$, $V$, and $I$ the total number of loops, vertices, and internal lines
respectively. 
\end{prob}

\begin{figure}[htb]
\bigskip
\centerline{\includegraphics*[width=11cm,angle=0]{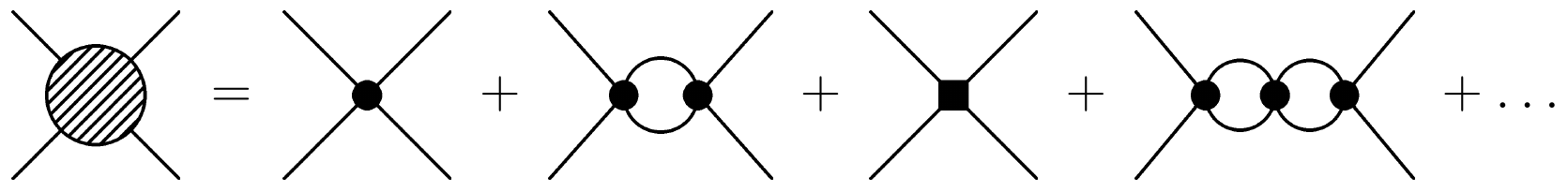}}
\medskip
\caption
{Diagrammatic expression for the two-body scattering amplitude $T_2$.
The circle (square) denotes a $C_0$ ($C_2$) interaction, respectively.}
\label{fig:amp2}
\end{figure}

The contact interactions in Eq.~(\ref{L-2body}) are ill-defined unless an
ultraviolet cutoff is imposed on the momenta in loop diagrams.
This can be seen by writing down the off-shell amplitude for two-body
scattering at energy $E$ in the center-of-mass frame at second order in 
perturbation theory:
\begin{equation}
 T_2 (E) \approx {-}C_0 - \frac\ii2
 C_0^2 \int \frac{\dd^3q}{(2 \pi)^3}
 \int\frac{\dd q_0}{2 \pi} \frac1{q_0 - q^2 /2m + \ii \epsilon}
 \frac1{E - q_0 - q^2 /2m + \ii\epsilon} + \cdots \,.
\nonumber\\
\label{A-pert}
\end{equation}
The two terms correspond to the first two diagrams in 
Fig.~\ref{fig:amp2}. 
The intermediate lines have momenta $\pm {\bm q}$.
The integral over $q_0$ in Eq.~(\ref{A-pert}) is easily
evaluated using contour integration:
\begin{equation}
 T_2 (E) \approx {-}C_0 - \frac12 C_0^2 
 \int \frac{\dd^3q}{(2 \pi)^3} \frac1{E - q^2/m +\ii \epsilon} + \cdots \,.
\end{equation}
The integral over ${\bm q}$ diverges. It can be regularized by imposing
an ultraviolet
cutoff $|{\bm q}| < \Lambda$. Taking the limit $\Lambda \gg |E|^{1/2}$,
the amplitude reduces to \footnote{
If the calculation was carried out in a
frame in which the total momentum of the two scattering particles was
nonzero, the simple cutoff $|{\bm q}| < \Lambda$ would give a result that does
not respect Galilean invariance. To obtain a Galilean-invariant result 
requires either using a more sophisticated cutoff or else imposing
the cutoff $|{\bm q}| < \Lambda$ only after an
appropriate shift in the integration variable ${\bm q}$.}
\begin{equation}
 T_2 (E) \approx - C_0 + \frac{mC_0^2}{4 \pi^2} 
 \left(\Lambda - \frac{\pi}{2} \sqrt {-mE -\ii \epsilon} \right) + \cdots \,.
\label{amp2-2nd}
\end{equation}

The dependence on the ultraviolet cutoff $\Lambda$ can be consistently
eliminated by a perturbative renormalization procedure. A simple choice
is to eliminate the parameter $C_0$ in favor of the scattering length
$a$, which is given by Eq.~(\ref{eq-ere}):
\begin{equation}
 a \approx \frac{mC_0}{8 \pi} \left( 1- \frac{m C_0 \Lambda}{4\pi^2}
 + \cdots\right) \,.
\end{equation}
Inverting this expression to obtain $C_0$ as a function of $a$ we obtain
\begin{equation}
 C_0  \approx \frac{8 \pi a}{m} \left( 1+ \frac{2 a \Lambda}{\pi}
 +  \cdots \right) \,,
\end{equation}
where we have truncated at second order in $a$. Inserting the expression
for $C_0$ into Eq.~(\ref{amp2-2nd}) and expanding to second order in $a$,
we obtain the renormalized expression for the amplitude:
\begin{equation}
 T_2 (E) \approx {-}\frac{8 \pi a}{m} \left( 1 + a \sqrt{-mE - \ii 
 \epsilon} + \cdots \right) = {-}\frac{8 \pi a}{m} \left( 1 - \ii ak +
 \cdots \right) \,.
\label{A2pert}
\end{equation}
If we evaluate this at the on-shell point $E=k^2/m$ and insert it into
Eq.~(\ref{amp2-2nd}), we find that it reproduces the first two terms in the
expansion of the universal scattering amplitude in Eq.~(\ref{eq-ere}) 
in powers of $ka$. By calculating $T_2 (E)$ to higher order 
in perturbation theory, we can reproduce the low-momentum 
expansion of Eq.~(\ref{eq-ere}) to higher order in $ka$. At the next order,
the $C_2$ term will contribute at tree level while $C_0$ will contribute
at the two-loop level. Thus a perturbative
treatment of the EFT reproduces the low-momentum
expansion of the two-body scattering amplitude.
The perturbative approximation is valid only if the momentum satisfies
$k \ll 1/a$.

A more interesting case occurs when the scattering length
is large, but all other effective range coefficients are still
determined by the scale $\Mhi$: $k\sim 1/|a|\sim\Mlo \ll \Mhi \sim 1/r_e$.
This scenario
is able to support shallow bound states with binding momentum of order $1/a$
and is relevant to ultracold atoms close to a
Feshbach resonance and very low-energy nucleons.
The scaling of the operators is then modified to:
\begin{equation}
 C_0 \sim \frac{1}{m\Mlo}\,,\quad C_2 \sim \frac{1}{m\Mlo^2\Mhi} \,,
 \quad\text{and}\quad
 D_0 \sim \frac{1}{m\Mlo^4}\,.
\label{eq:scalingMlo}
\end{equation}
The factors of $\Mlo$ in amplitudes can now come from small momenta and from the
coupling constants.  Above we adjusted the scaling of the three-body coupling 
$D_0$ as well, foreclosing a result discussed below Eq.~\eqref{BhvK:general}.

With the scaling as in Eq.~\eqref{eq:scalingMlo}, the power counting expression 
in Eq.~(\ref{nu-pert}) is therefore modified to
\begin{equation}
 \nu=3L+2+\sum_i (i-3) V_{2i} \geq -1\,.
\end{equation}
If we are interested in two-body observables involving energy $E \sim 1/a^2$,
such as shallow bound or virtual states,
we must resum the diagrams involving only $C_0$ interactions
to all orders~\cite{vanKolck:1998bw,Kaplan:1998we}.  Without this resummation, 
our EFT would break down not at $\Mhi$, but already at the much smaller scale 
$1/|a|\sim\Mlo$.  In the scenario assumed here, all higher-derivative two-body 
interactions ($C_2$ and beyond) still involve inverse powers of $\Mhi$ and are 
thus perturbative.

\begin{figure}[htb]
\bigskip
\centerline{\includegraphics*[width=6.5cm,angle=0]{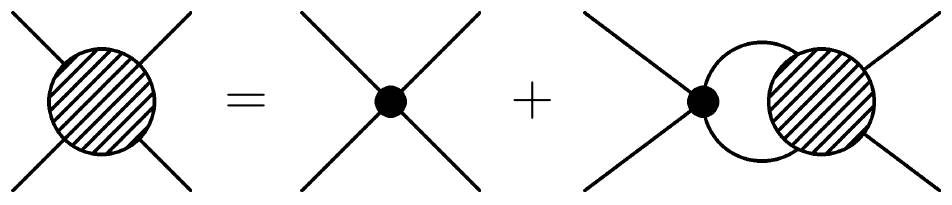}}
\medskip
\caption
    {Integral equation  for the two-body scattering amplitude $T_2$
      at leading order in the case of large scattering length.
      Notation as in Fig.~\ref{fig:amp2}.}
\label{fig:amp2b}
\end{figure}

The resummation of $C_0$ interactions is most easily accomplished by
realizing that
the corresponding Feynman diagrams in Fig.~\ref{fig:amp2}
form a geometric series. 
Summing the geometric series, the exact expression
for the amplitude is
\begin{equation}
 T_2(E) = {-}C_0 \left[ 1 + \frac{mC_0}{4 \pi^2}
 \left( \Lambda - \frac{\pi}{2} \sqrt {-mE -\ii\epsilon} \right) 
 \right]^{-1} \,.
\label{A-nonpert}
\end{equation}
Alternatively, we can use the fact that summing the $C_0$ diagrams in
Fig.~\ref{fig:amp2} is equivalent to solving the following
integral equation:
\begin{equation}
 T_2(E) = {-}C_0 - \frac{\ii}{2} C_0 \int \frac{\dd^3q}{(2\pi)^3}
 \int\frac{\dd q_0}{2\pi} \frac1{q_0 - q^2/2m + \ii \epsilon}
 \frac1{ E - q_0 - q^2/2m + \ii \epsilon} \, T_2 (E) \,.
\label{inteq-2}
\end{equation}
The integral equation is expressed diagrammatically in Fig.~\ref{fig:amp2b}.
Since the function $T_2(E)$ is independent of ${\bm q}$ and $q_0$,
it can be pulled
outside of the integral in  Eq.~(\ref{inteq-2}). The integral can be
regularized by imposing an ultraviolet cutoff $\Lambda$.
The integral equation is now trivial to solve and the
solution is given in Eq.~(\ref{A-nonpert}).

\begin{figure}[htb]
\bigskip
\centerline{\includegraphics*[width=12cm,angle=0]{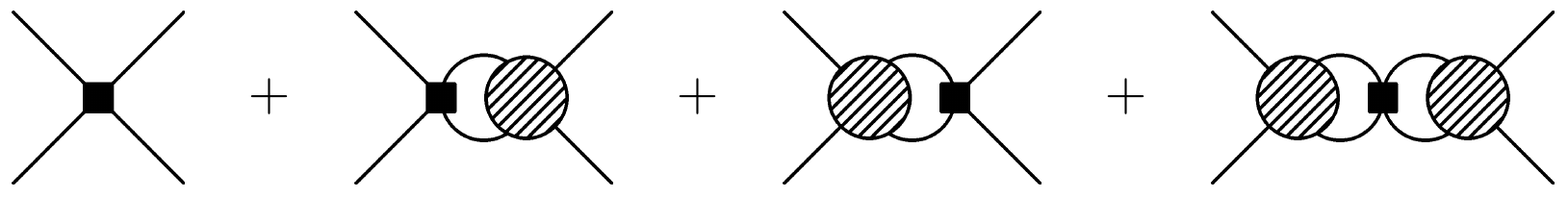}}
\medskip
\caption
    {Next-to-leading order correction to $T_2$.
      Notation as in Fig.~\ref{fig:amp2}.}
\label{fig:amp2c}
\end{figure}

The expression for the resummed two-body
amplitude in Eq.~(\ref{A-nonpert}) depends on the
parameter $C_0$ in the Lagrangian and
on the ultraviolet cutoff $\Lambda$. As in the perturbative case,
renormalization can be implemented
by eliminating $C_0$ in favor of a low-energy observable, such as the
scattering length $a$. 
Matching the resummed expression to the effective range expansion
for $T_2$, we obtain
\begin{equation}
 C_0 = \frac{8\pi a}{m} \left( 1 - \frac{2 a \Lambda}{\pi} \right)^{-1}  \,.
\label{g2-tune}
\end{equation}
Given a fixed ultraviolet cutoff $\Lambda$, this equation prescribes how
the parameter $C_0$ must be tuned in order to give the 
scattering length $a$.  Note that for $\Lambda \gg 1/|a|$,
the coupling constant $C_0$ is always negative 
regardless of the sign of $a$.
Eliminating $C_0$ in Eq.~(\ref{A-nonpert}) in favor of $a$, 
we find that the resummed amplitude reduces to
\begin{equation}
 T_2 (E) = \frac{8\pi}{m}\frac1{{-}1/a + \sqrt {{-}mE -\ii \epsilon}} \,,
\label{a-npren}
\end{equation}
which reproduces the effective range expansion of the scattering
amplitude by construction. 
In this simple case, we find that our renormalization prescription 
eliminates the dependence on $\Lambda$ completely. In general, 
we should expect it to only be suppressed by powers of
$1/(a \Lambda)$ or $mE/ \Lambda^2$.  A final step of taking the limit 
$\Lambda \to \infty$ would then be required to obtain results 
that are completely independent of  $\Lambda$.
The first correction to Eq.~(\ref{a-npren}) is given by the
$C_2$ interaction. The corresponding diagrams are shown in
Fig.~\ref{fig:amp2c}.
After the matching, the final result for the scattering amplitude
at next-to-leading order is
\begin{equation}
 T_2 (E) = \frac{8\pi}{m}\left(
 \frac1{{-}1/a + \sqrt {{-}mE -\ii\epsilon}}
 + \frac{r_e mE/2}{({-}1/a + \sqrt {-mE - \ii \epsilon})^2}\right) \,,
\label{a-npren2}
\end{equation}
where $r_e$ is the effective range. 
The derivation of this expression will be left as an exercise.

\begin{prob}
{\emph Exercise:} 
Derive the  next-to-leading order correction in Eq.~(\ref{a-npren2})
by calculating the loop diagrams in Fig.~\ref{fig:amp2c}. Neglect 
all terms that vanish as $\Lambda \to \infty$. Introduce a 
next-to-leading order piece of $C_0$ to cancel the cubic divergence. 
\end{prob}

\subsection{Dimer field formalism}

In applications to systems with more than two particles, it is often useful to 
rewrite the EFT for short-range interactions specified by the 
Lagrangian~(\ref{L-2body}) using so-called dimer fields
$d$~\cite{Kaplan:1996nv}:
\begin{spliteq}
 {\mathcal L} &= \phi^\dagger \left( \ii \partial_t + \frac{\nabla^2}{2m} 
 \right) \phi +{g_0} d^\dagger d
 +g_2  d^\dagger\left( \ii \partial_t + \frac{\nabla^2}{4m} 
 \right)d+\cdots \\
 &\phantom{=}- {y} \left( d^\dagger \phi^2 + {\phi^\dagger}^2 d \right)
 -{d_0} d^\dagger d \phi^\dagger \phi +\cdots\,.
\label{L-BHvK}
\end{spliteq}
One important feature of this Lagrangian is that there is no direct two-body 
contact interaction term $(\phi^\dagger \phi)^2$.  All interactions between 
$\phi$ particles are mediated via exchange of a dimer field $d$, \ie, we have 
effectively performed a Hubbard--Stratonovich transformation.  Eliminating 
the dimer field $d$ by using its equations of motion, it can be shown that the 
physics of this EFT is equivalent to the Lagrangian~(\ref{L-2body}).

Note that the Lagrangian~(\ref{L-BHvK}) contains one more free parameter than 
the Lagrangian (\ref{L-2body}).  Thus some parameters are redundant.  For the 
leading-order case ($g_n=0$ for $n\geq 2$), \eg, we find explicitly,
\begin{equation}
 C_0=\frac{4y^2}{g_0}\,,
\label{eq-relation}
\end{equation}
such that $y$ and $g_0$ are not independent.  Higher-order corrections can be 
obtained by including a kinetic-energy term for the dimer field.  The constants 
$g_0$ and $y$ then become independent and can be related to combinations of 
$C_0$ and $C_2$ in the theory without dimers.  Here, we only discuss the 
leading-order case. 

\begin{prob}
{\emph Exercise:} Derive Eq.~(\ref{eq-relation}) using the classical 
equation of motion for $d$.
\end{prob}

\begin{figure}[htb]
\bigskip
\centerline{\includegraphics*[width=12cm,angle=0]{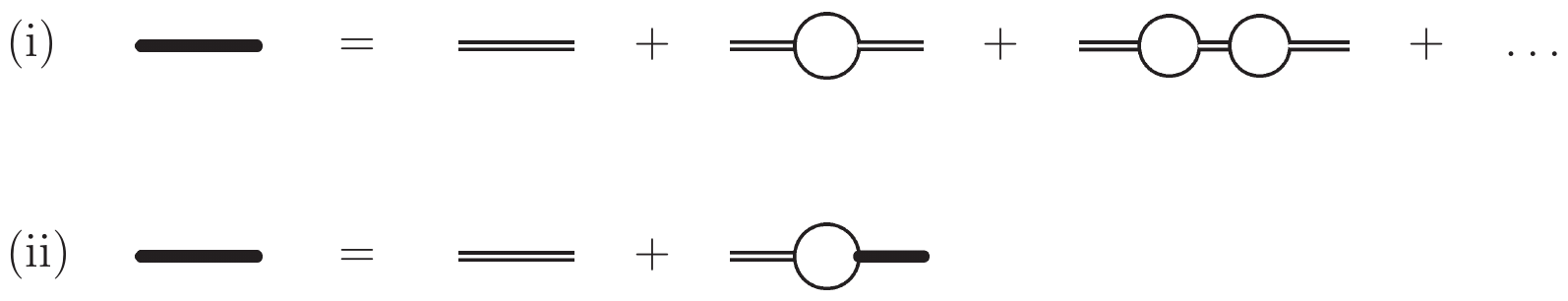}}
\medskip
\caption
{Diagrammatic equations for the full dimer propagator
$i D(P_0,P)$. Thin (thick)
solid lines represent particle (full dimer) propagators.
Double lines indicate bare dimer propagators.
(i) perturbative expansion in powers of $y$,
(ii) integral equation summing the geometric series in (i).}
\label{fig:bubbles}
\end{figure}

The bare propagator for the dimer field is simply the constant $\ii/g_0$,
which corresponds to no propagation in space or time.  However, there are 
corrections to the dimer propagator from the diagrams in 
Fig.~\ref{fig:bubbles}(i) which allow the dimer to propagate.  This is 
completely analogous to the geometric series we found we had to sum to obtain 
the leading-order scattering amplitude.  In Feynman diagrams, we represent the 
full dimer propagator $\ii D(P_0,P)$ by a thick solid line.  We can 
calculate the full dimer propagator by solving the simple integral equation 
shown in Fig.~\ref{fig:bubbles}(ii). The loop on the right side is just the 
integral in Eq.~(\ref{A-pert}), with $E$ replaced by $P_0-P^2/(4m)$, where $P_0$ 
and ${\bm P}$ are the energy and momentum of the dimer.  The solution for the 
full dimer propagator is
\begin{equation}
 \ii D(P_0,P) = \frac{2\pi i}{y^2 m}
 \left[ \frac{2\pi g_0}{y^2 m} + \frac{2}{\pi} \Lambda -
  \sqrt{{-}mP_0 + P^2/4 - \ii \epsilon} \right]^{-1} \,,
\label{diprop}
\end{equation}
where as before $\Lambda$ is a cutoff on the loop momentum in the bubbles.
Using Eq.~(\ref{eq-relation}) and making the substitution given in 
Eq.~(\ref{g2-tune}), the expression for the complete dimer propagator is
\begin{equation}
 \ii D(P_0, P) = -\frac{2\pi \ii}{y^2 m}
 \left[ {-}1/a + \sqrt{{-}mP_0 + P^2/4 -\ii \epsilon}
 \right]^{-1} .
\label{propdiatom}
\end{equation}
Note that all the dependence on the ultraviolet cutoff is now in the
multiplicative factor $1/y^2$.  The complete dimer propagator differs from the 
off-shell two-body amplitude $T_2$ in Eq.~(\ref{a-npren}) only by a 
multiplicative constant.  For $a>0$, it has a pole at $P_0 = -1/(ma^2) + P^2/4$ 
corresponding to a dimer of momentum ${\bm P}$ and binding energy 
$B_2=1/(ma^2)$.  As $P_0$ approaches the dimer pole, the limiting behavior of 
the propagator is
\begin{equation}
 D(P_0, P) \longrightarrow
 \frac{Z_D}{P_0 - ({-}1/(ma^2)+ P^2/4) + \ii \epsilon} \,,
\label{dimer-pole}
\end{equation}
where the residue factor is
\begin{equation}
 Z_D= \frac{4\pi}{a m^2 y^2} \,.
\label{dimer-Z}
\end{equation}
If we regard the composite operator $d$ as a quantum field that annihilates and 
creates dimers, then $Z_D$ is the wave function renormalization constant for 
that field.  The renormalized propagator $Z_D^{-1} D(P_0,P)$ is completely 
independent of the ultraviolet cutoff.

\subsection{Three-body system}
\label{sec:EFT-ThreeBosons}

\begin{figure}[htb]
\bigskip
\centerline{\includegraphics*[width=9cm,angle=0]{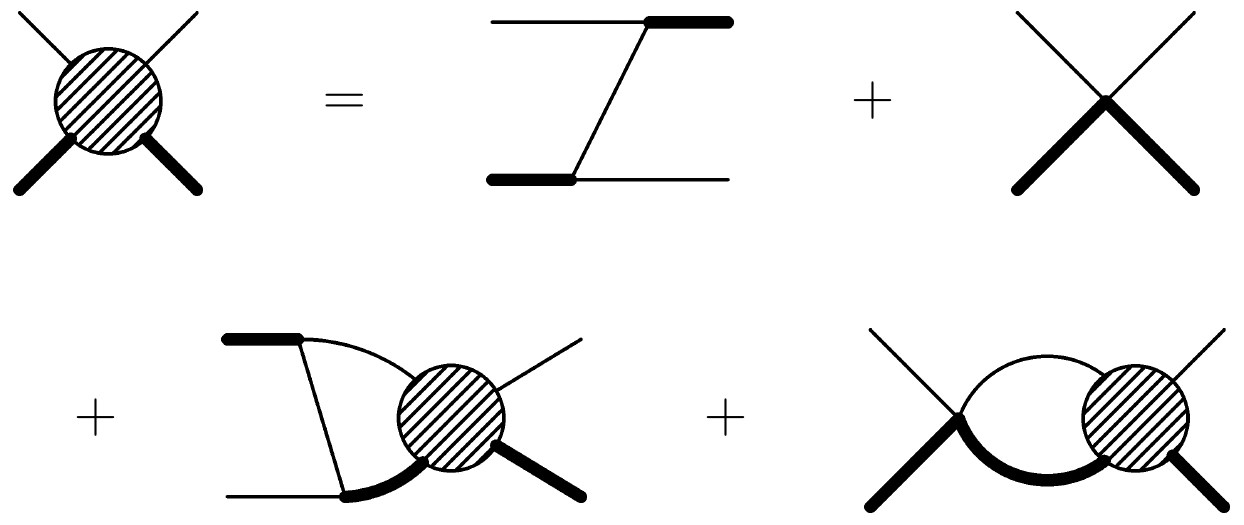}}
\medskip
\caption
{The integral equation for the three-body amplitude $T_3$. Thin (thick)
solid lines represent particle (full dimer) propagators. External lines
are amputated.}
\label{fig:inteq12}
\end{figure}

We now study the amplitude for particle-dimer scattering $T_3$.  The simplest 
diagram we can write down involving only two-body interactions is the exchange 
of a particle between in- and outgoing dimers.  With the scaling of low-energy 
constants as in Eq.~\eqref{eq:scalingMlo}, the power counting implies that all 
diagrams that are chains of such exchanges are equally important, \ie, they have 
to be summed up nonperturbatively.  Just like in the two-body case, this can be 
written as an integral equation.  Also including the three-body interaction 
(note $D_0 \to d_0$ in the Lagrangian with dimer fields), we get the result that 
is shown diagrammatically in Fig.~\ref{fig:inteq12}.\footnote{Note that this 
amplitude is well defined even if $a<0$ and there is no two-body bound state. 
In this case particle lines must be attached to the external dimer propagators 
to obtain the 3-particle scattering amplitude.}

Omitting the three-body interaction, 
this is exactly the well-known Skorniakov-Ter-Martirosian (STM)
integral equation~\cite{Skorniakov:1957aa}, which the EFT with
Lagrangian~(\ref{L-BHvK}) reproduces by construction.  In addition, EFT 
provides a clear method to renormalize this equation with a three-body 
interaction and thus remove its pathologies (discussed below).

In Fig.~\ref{fig:inteq12}, all external lines are understood to be amputated.
It simply gives the non-perturbative solution of the three-body problem
for the interaction terms proportional to $g_0$, $y$, and $d_0$ in 
Eq.~(\ref{L-BHvK}).

The two tree diagrams on the right side of Fig.~\ref{fig:inteq12} constitute the 
inhomogeneous term in the integral equation.  An iterative ansatz for the 
solution of this equation shows that all diagrams with the $g_0$, $y$, and 
$d_0$ interactions are generated by the iteration.  Note also that the thick 
black lines in Fig.~\ref{fig:inteq12} represent the full dimer propagator given 
in Eq.~(\ref{propdiatom}).

In the center-of-mass frame, we can take the external momenta of the
particle and dimer to be $-{\bm p}$ and $+{\bm p}$ for the incoming lines
and $-{\bm k}$ and $+{\bm k}$ for the outgoing lines.  We take their
energies to be $E_A$ and $E-E_A$ for the incoming lines and $E_A'$ and
$E-E_A'$ for the outgoing lines. The amplitude $T_3$ is then a
function of the momenta ${\bm p}$ and ${\bm k}$ and the energies $E$,
$E_A$ and $E_A'$. The integral equation involves a loop over the
momentum $-{\bm q}$ and energy ${q_0}$ of a virtual particle. Using the
Feynman rules encoded in the Lagrangian (\ref{L-BHvK}), we obtain
\begin{multline}
 T_3 ({\bm p}, {\bm k}; E, E_A, E_A')
 = {-}\left[ \frac{4y^2}{ E-E_A-E_A' - ({\bm p} + {\bm k})^2 /(2m) + \ii
 \epsilon}
 + {d_0} \right] \\
 \null + \frac{2\pi \ii}{m y^2} \int \frac{\dd q_0}{2 \pi}
 \int\frac{\dd^3q}{(2 \pi)^3}
 \left[ \frac{4 y^2}{E-E_A - q_0 - ({\bm p} + {\bm q})^2/(2m) + \ii\epsilon}
 + {d_0} \right] \\
 \times \frac1{q_0 - q^2/(2m) + \ii \epsilon}\;
 \frac{T_3 ({\bm q}, {\bm k}; E, q_0, E_A')}
 {1/a - \sqrt{-m(E-q_0) + q^2 /4 -\ii \epsilon} } \,.
\end{multline}
The integral over $q_0$ can be evaluated by contour integration. This
sets $q_0 = q^2/(2m)$, so the amplitude $T_3$ inside the integral has the
incoming particle on-shell. 

We obtain a simpler integral
equation if we also set the energies of both the initial and final particles in
$T_3$ on-shell: $E_A = p^2/(2m)$, $E_A' = k^2/(2m)$.
Thus only the dimer lines have energies that are off-shell.
The resulting integral equation is
\begin{multline}
 T_3 \left({\bm p}, {\bm k}; E, \frac{p^2}{2m}, \frac{k^2}{2m}\right)
 = {-}{4my^2} 
 \left[ \frac1{mE - (p^2 + {\bm p} \cdot {\bm k} + k^2) + \ii\epsilon} 
 + \frac{d_0}{4 my^2} \right] \\
 \null - 8 \pi \int\frac{\dd^3q}{(2 \pi)^3}
 \left[ \frac1{mE - (p^2 + {\bm p} \cdot {\bm q} + q^2) + \ii \epsilon}
 + \frac{d_0}{4my^2} \right] \\
 \null\times
 \frac{T_3 ({\bm q}, {\bm k}; E, q^2/(2m), k^2/(2m))}
 {{-}1/a + \sqrt{{-}mE + 3q^2 /4 -\ii \epsilon} } \,.
\label{BhvK:general}
\end{multline}
This is an integral equation with three integration variables 
for an amplitude $T_3$ that depends explicitly 
on seven independent variables.  There is also an additional implicit variable
provided by an  ultraviolet cutoff $|{\bm q}| < \Lambda$
on the loop momentum.

If we set $d_0 = 0$ and ignore the ultraviolet cutoff, 
the integral equation in Eq.~(\ref{BhvK:general}) is equivalent to the 
Skorniakov-Ter-Martirosian (STM) equation, 
an integral equation for three particles interacting via
zero-range two-body forces derived by Skorniakov and Ter-Martirosian 
in 1957~\cite{Skorniakov:1957aa}.  
It was shown by Danilov that the STM equation has no unique 
solution in the case of identical bosons~\cite{Danilov:1961aa}. He also
pointed out that a unique solution could be obtained if one 
three-body binding energy is fixed. 
Kharchenko was the first to solve the STM equation with a finite 
ultraviolet cutoff that was tuned to fit observed three-body data.  
Thus the cutoff was treated 
as an additional parameter~\cite{Kharchenko:1973aa}. 
When we discuss the running of $d_0$, we will see
that this \adhoc procedure is indeed justified and emerges
naturally when the three-body equation is renormalized~\cite{Bedaque:1998kg}.

Here we restrict our attention to the sector of the three-body problem with 
total orbital angular momentum $L=0$ where the three-body interaction
contributes. For higher $L$, the original STM equation has a unique solution
and can be solved numerically without complication.

The projection onto $L=0$ can be accomplished
by averaging the integral equation over the cosine
of the angle between ${\bm p}$ and ${\bm k}$: $x={\bm p}\cdot{\bm k}/
(pk)$. It is also convenient to multiply the amplitude $T_3$
by the wave function renormalization factor $Z_D$ given in 
Eq.~(\ref{dimer-Z}).
We will denote the resulting amplitude by $T_3^0$:
\begin{equation}
 T_3^0(p, k; E) \equiv Z_D 
 \int_{-1}^1 \! \frac{\dd x}{2}\,
 T_3 \left({\bm p}, {\bm k}; E, p^2/(2m), k^2/(2m)\right) .
\label{A-def}
\end{equation}
Furthermore, it is convenient to express the three-body coupling constant 
in the form
\begin{eqnarray}
 d_0 = {-}\frac{4my^2}{\Lambda^2} H(\Lambda) \,.
\label{g3g2}
\end{eqnarray}
Since $H$ is dimensionless, it can only  be a
function of the dimensionless variables $a \Lambda$ and 
$\Lambda/ \Lambda_*$, where $\Lambda_*$ is a three-body 
parameter defined below. We will find that $H$ is a function of
$\Lambda/ \Lambda_*$ only.  

The resulting integral equation is:
\begin{multline}
 T_3^0 (p, k; E)  = \frac{16 \pi}{m a} 
 \left[ \frac{1}{2pk} \ln \left(\frac{p^2 + pk + k^2 -mE - \ii \epsilon}
 {p^2 - pk + k^2 - mE - \ii \epsilon}\right) + \frac{H(\Lambda)}{\Lambda^2} 
 \right] \\
 + \frac{4}{\pi} \int_0^\Lambda \dd q \, q^2
 \left[\frac{1}{2pq} \ln\left( \frac{p^2 +pq + q^2 - mE - \ii \epsilon}
 {p^2 - pq + q^2 -mE -\ii \epsilon}\right)
 + \frac{H(\Lambda)}{\Lambda^2} \right] \\
 \null \times \frac{ T_3^0 (q, k; E)}
 {{-}1/a + \sqrt{3q^2/4 -mE - \ii\epsilon}} \,.
\label{BHvK}
\end{multline}
Note that the ultraviolet cutoff $\Lambda$ on the
integral over $q$ has been made explicit. 
A change in the endpoint $\Lambda$ of the loop integral
should be compensated by the $\Lambda$-dependence of the function 
$H$ in Eq.~(\ref{BHvK}).
More specifically, $H$ must be tuned as a function of $\Lambda$ 
so that the cutoff dependence of the solution $T_3^0 (p, k; E)$
of Eq.~(\ref{BHvK}) decreases as a power of $\Lambda$.  This will 
guarantee that $T_3^0 (p, k; E)$ has a well-behaved limit 
as $\Lambda \to \infty$. The renormalization group behavior of
$H$ will be discussed in detail below. In the next subsection, we 
show how
different three-body observables can be obtained from the 
solution $T_3^0 (p, k; E)$ of Eq.~(\ref{BHvK}).

\begin{prob}
Fill in the gaps in the above derivation of Eq.~(\ref{BHvK}) and
generalize the derivation to general angular momentum $L$.
\end{prob}


\subsection{Three-body observables}
\label{sec:EFT-3BObs}

The solution $T_3^0 (p, k; E)$ to the three-body
integral equation~(\ref{BHvK})
encodes all information about three-body observables in the sector with total
orbital angular momentum quantum number $L=0$.
In particular, it contains information about the 
binding energies $B_3^{(n)}$ of the three-body bound 
states~\cite{Efimov:1970aa}.
For a given ultraviolet cutoff $\Lambda$, the amplitude 
$T_3^0 (p, k; E)$ has a finite number of poles in $E$
corresponding to the bound states whose binding energies
are less than about $\Lambda^2$.  As $\Lambda$ increases,
new poles emerge corresponding to deeper bound states.
In the limit $\Lambda \to \infty$, the locations of these poles 
approach the energies $-B_3^{(n)}$ of the three-body bound states.
The residues of the poles of $T_3^0 (p, k; E)$
factor into functions of $p$ and functions of $k$:
\begin{equation}
 T_3^0 (p, k; E) \longrightarrow 
 \frac{ {\mathcal B}^{(n)}(p) {\mathcal B}^{(n)}(k)}{ E + B_3^{(n)} } \,,
 \qquad \text{as}\ E \to {-}B_3^{(n)} \,.
\end{equation}
Matching the residues of the poles on both sides of Eq.~(\ref{BHvK}),
we obtain the bound-state equation
\begin{spliteq}
 {\mathcal B}^{(n)}(p) &=  \frac{4}{\pi} \int_0^\Lambda \! \dd q \, q^2
 \left[\frac{1}{2pq} \ln \frac{p^2 +pq + q^2 - mE - \ii \epsilon}
 {p^2 - pq + q^2 -mE -\ii \epsilon} + \frac{H(\Lambda)}{\Lambda^2} \right] \\
 &\phantom{=}\times \left[{-}1/a + \sqrt{3q^2/4 -mE
 -\ii \epsilon} \right]^{-1} {\mathcal B}^{(n)}(q) \,.
\label{BHvK-homo}  
\end{spliteq}
The values of $E$ for which this homogeneous integral equation has 
solutions are the energies ${-}B_3^{(n)}$ of the three-body states.
For a finite ultraviolet cutoff $\Lambda$,
the spectrum of $B_3^{(n)}$ is cut off around $\Lambda^2$.

\begin{figure}[htb]
\bigskip
\centerline{\includegraphics*[width=11cm,angle=0]{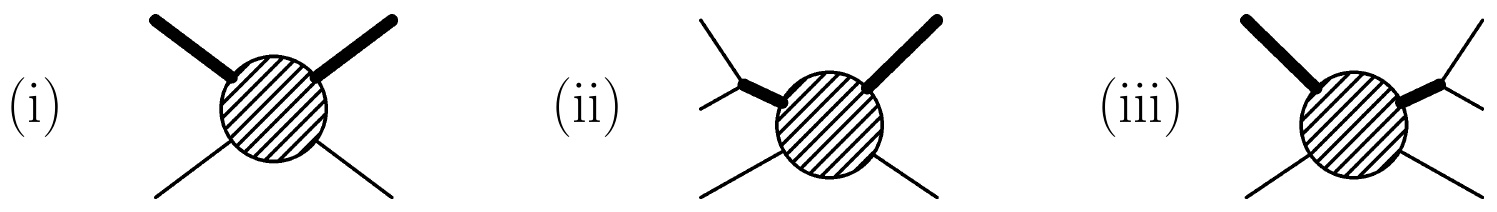}}
\medskip
\caption
{Amplitudes for (i) particle-dimer scattering, (ii) three-body recombination, 
and (iii) three-body breakup. 
Diagrams (ii) [(iii)] should be summed over the three pairs of 
particles that can interact first [last].  
Notation as in Fig.~\ref{fig:inteq12}.}
\label{fig:3br}
\end{figure}

The S-wave phase shifts for particle-dimer scattering can be determined
from the solution $T_3^0 (p, k; E)$ to the 
integral equation~(\ref{BHvK}).
The T-matrix element for the elastic scattering of an particle and a dimer 
with momenta $k$ is given by the amplitude $T_3^0$ 
evaluated at the on-shell point $p=k$ and $E= {-}B_2 + 3k^2/(4m)$
and multiplied by a wave function renormalization factor $Z_D^{1/2}$ 
for each dimer in the initial or final state.
It can be represented by the Feynman diagram in 
Fig.~\ref{fig:3br}(i).  The blob represents the amplitude $T_3$ or
equivalently $Z_D^{-1} T_3^0$.
The external double lines are amputated and correspond 
to asymptotic dimers and are associated with factors $Z_D^{1/2}$

The S-wave contribution to the T-matrix element is
\begin{equation}
 T_{PD \to PD}^{0} = 
 T_3^0 (k, k; 3k^2/(4m)-1/(ma^2)) \,,
\end{equation}
where $B_2=1/(ma^2)$ has been used.
Note that the factors of $Z_D$ multiplying $T_3^0$ cancel.
The differential cross section for elastic particle-dimer scattering is
\begin{equation}
 \dd \sigma_{PD \to PD}
 = \frac{2 m}{3 k} \left| T_{PD \to PD}(k) \right|^2
 \frac{k m}{6 \pi^2} d \Omega \,.
\label{T-AD}
\end{equation}
The flux factor $2m/(3 k)$ is the inverse of the relative velocity 
of the particle and the dimer.  The phase space factor 
$k md \Omega/(6 \pi^2)$ takes into account energy and momentum 
conservation and the standard normalization of momentum eigenstates:
\begin{equation}
 \int \frac{\dd^3 p_A}{(2 \pi)^3} \frac{\dd^3 p_D}{(2 \pi)^3} 
 (2 \pi)^4 \delta^3({\bm p}_A + {\bm p}_D)
 \delta(p_A^2/(2m) + p_D^2/(4m) - E)
 = \frac{m}{6 \pi^2} (4 mE/3)^{1/2} \int \dd\Omega \,.
\end{equation}
The S-wave phase shift for particle-dimer scattering
is related to the T-matrix element via
\begin{equation}
 \frac1{k\cot\delta^{PD}_0 (k) -\ii k}
 = \frac{m}{3 \pi} T_3^0 (k, k; 3k^2/(4m)-1/(ma^2)) \,.
\label{T12}
\end{equation}
In particular, the particle-dimer scattering length is given by
\begin{equation}
 a_{PD} = {-}\frac{m}{3 \pi} T_3^0 (0, 0; {-}1/(ma^2)) \,.
\end{equation}

The threshold rate for three-body recombination can also be obtained 
from the solution $T_3^0 (p, k; E)$ to the three-body integral 
equation in Eq.~(\ref{BHvK}). This is possible only at threshold, 
because a 3-particle scattering state becomes pure $L=0$ only
in the limit that the energies of the particles go to zero. 
The T-matrix element for the recombination process can be represented 
by the Feynman diagram in Fig.~\ref{fig:3br}(ii)
summed over the three pairs of particle lines that can attach 
to the dimer line.
The blob represents the amplitude $Z_D^{-1} T_3^0$ evaluated 
at the on-shell point $p=0$ $k=2/(\sqrt{3}\, a)$, and $E=0$.
The solid line represents the dimer propagator $\ii D(0,0)$
evaluated at zero energy and momentum $2/(\sqrt{3}\, a)$, which is given by 
Eq.~(\ref{diprop}).  The factor for the particle-dimer vertex is $-\ii 2y$.
The wave function renormalization factor $Z_D^{1/2}$ for the
final-state dimer is given by Eq.~(\ref{dimer-Z}).
In the product of factors multiplying $T_3^0$, 
the dependence on $y$ and $\Lambda$ can be eliminated in favor 
of the scattering length $a$.
Taking into account a factor of 3 from the three Feynman diagrams,
the T-matrix element is
\begin{equation}
 T_{PD \to PPP} = 6\sqrt{\pi a^3}\, 
 T_3^0 (0, 2/(\sqrt{3}a);0) \,.
\end{equation}
The differential rate $dR$ for the recombination of three particles
with energies small compared to the dimer binding energy
can be expressed as
\begin{equation}
 \dd R = \left| T_{PPP \to PD} \right|^2
 \frac{k m}{6 \pi^2}\, \dd \Omega \,,
\label{dR-T}
\end{equation}
where $k = 2/(\sqrt{3} a)$.
The threshold rate for three-body breakup can be obtained
in a similar way from  the Feynman diagram in Fig.~\ref{fig:3br}(iii)

The inhomogeneous integral equation for the off-shell particle dimer 
amplitude, Eq.~(\ref{BHvK}), and the homogeneous equation, 
Eq.~(\ref{BHvK-homo}), for the three-body binding energies afford
no analytical solution. They are usually solved by 
discretizing the integrals involved and solving the resulting
matrix problems numerically.

\subsection{Renormalization group limit cycle}
\label{sec:RGlc}

The form of the full renormalized dimer propagator in
Eq.~(\ref{propdiatom}) is consistent with the continuous scaling symmetry
\begin{equation}
 a \longrightarrow \lambda a \,,
 \qquad
 E \longrightarrow \lambda^{-2} E \,,
\label{scaling-1}
\end{equation}
for any positive real number $\lambda$.
In the integral equation~(\ref{BHvK}), this scaling
symmetry is broken by the ultraviolet cutoff on the integral and by the
three-body terms proportional to $H/ \Lambda^2$. To see that the cutoff and
the three-body terms are essential, we set $H=0$ and take
$\Lambda \rightarrow \infty$. The resulting integral equation has exact
scaling symmetry. We should therefore expect its solution 
$T_3^0 (p,k; E)$ to behave asymptotically as $p \rightarrow \infty$ 
like a pure power of $p$. Neglecting the inhomogeneous term,
neglecting $E$ and $1/a^2$ compared to $q^2$,
and setting $T_3^0 \approx p^{s-1}$, 
the integral equation reduces to~\cite{Danilov:1961aa}
\begin{equation}
 p^{s-1} = \frac{4}{\sqrt{3} \pi p} \int_0^\infty \dd q \, q^{s-1} 
 \ln \frac{p^2 + pq + q^2}{p^2 -pq + q^2} \,.
\end{equation}
Making the change of variables
$q = xp$, the dependence on $p$ drops out, and we obtain
\begin{equation}
 1 = \frac{4}{\sqrt{3} \pi} \int_0^\infty \dd x \, x^{s-1} 
 \ln \frac{1 + x + x^2}{1 -x + x^2} \,.
\end{equation}
The integral is a Mellin transform that can be evaluated
analytically.  The resulting equation for $s$ is
\begin{equation}
 1 = \frac{8}{\sqrt{3} s} \frac{\sin (\pi s/6)}{\cos (\pi s/2)} \,.
\end{equation}
The solutions with the lowest values of $|s|$ are
purely imaginary: $s = \pm \ii s_0$, where $s_0 \approx 1.00624$. The most
general asymptotic solution therefore has two arbitrary constants:
\begin{equation}
 T_3^0 (p, k; E) \longrightarrow A_+ \, 
 p^{-1+\ii s_0} + A_- \, p^{-1-\ii s_0} 
 \,,\qquad \text{as}\ p \to \infty \,.
\end{equation}
The inhomogeneous term in the integral equation~(\ref{BHvK}) will
determine one of the constants. The role of the three-body term in the
integral equation is to determine the other constant, thus giving the
integral equation a unique solution.

By demanding that the solution of the integral equation~(\ref{BHvK}) 
has a well-defined limit as $\Lambda \to \infty$, 
Bedaque~\etal deduced the $\Lambda$-dependence 
of $H$ and therefore of $d_0$~\cite{Bedaque:1998kg}. 
The leading dependence on $\Lambda$ on the right side of the
three-body integral equation in Eq.~(\ref{BHvK}) as $\Lambda \to \infty$ 
is a log-periodic term of order $\Lambda^0$ that comes from the 
region $q \sim \Lambda$.  
There are also contributions of order $1/\Lambda$
from the region $|a|^{-1},k,|E|^{1/2} \ll q \ll \Lambda$,
which have the form
\begin{equation}
 \frac{8}{\pi\sqrt{3}} \int^\Lambda \dd q \, 
 \left( \frac{1}{q^2}+ \frac{H(\Lambda)}{\Lambda^2}\right)
 (A_+ \, q^{+\ii s_0} + A_- \, q^{-\ii s_0}) \,.
\label{1overLambda}
\end{equation}
The sum of the two terms will decrease even faster as $1/\Lambda^2$ 
if we choose the function $H$ to have the form
\begin{equation}
 H(\Lambda) =
 \frac{ A_+ \Lambda^{\ii s_0}/(1-\ii s_0) + A_- \Lambda^{-\ii s_0}/(1+\ii s_0)}
 {A_+ \Lambda^{\ii s_0}/(1+\ii s_0)  + A_- \Lambda^{-\ii s_0}/(1-\ii s_0)} \,.
\label{H-tune}
\end{equation}
The tuning of $H$ 
that makes the term in Eq.~(\ref{1overLambda}) decrease like
$1/\Lambda^2$ also suppresses the contribution from the region
$q \sim \Lambda$ by a power of $1/\Lambda$  so that it goes to 0 
in the limit $\Lambda \to \infty$.
By choosing $A_\pm = (1 + s_0^2)^{1/2} \Lambda_*^{\mp \ii s_0}/2$
in Eq.~(\ref{H-tune}), we obtain~\cite{Bedaque:1998kg}
\begin{equation}
 H (\Lambda) \approx \frac{\cos [s_0 \ln (\Lambda/ \Lambda_*) + \arctan s_0]}
 {\cos [s_0 \ln (\Lambda/ \Lambda_*) - \arctan s_0]} \,.
\label{H-Lambda}
\end{equation}
This equation defines a three-body scaling-violation parameter 
$\Lambda_*$ with dimensions of momentum. The value of $\Lambda_*$
can be fixed from a three-body datum. All other three-body
observables can then be predicted. If Eq.~(\ref{H-Lambda}) is 
substituted back into the three-body equation~(\ref{BHvK})
for numerical calculations, it must be multiplied by a normalization factor
$b\approx 1$ whose precise value
depends on the details of the regularization~\cite{Braaten:2011sz}.

Note that $H$ is a $\pi$-periodic function of 
$s_0\ln(\Lambda/\Lambda_*)$, so $\Lambda_*$ is defined only up to a 
multiplicative factor of $(\eex^{\pi/s_0})^n$, where $n$ is an integer.
Thus the scaling symmetry of Eq.~(\ref{scaling-1}) is broken to
the discrete subgroup of scaling transformations with multiples of
the preferred scaling factor $\lambda=\eex^{\pi/s_0}$. This discrete 
scaling symmetry is, \eg, evident in the geometric 
spectrum of three-body Efimov states~\cite{Efimov:1970aa} 
in the unitary limit ($1/a=0$) that naturally emerge in this EFT:
\begin{equation}
 B_3^{(n)}\approx 0.15 \lambda^{2(n_*-n)}\,\frac{\Lambda_*^2}{m}\,,
\end{equation}
where $n_*$ an integer labeling the state
with binding energy  
$0.15\,\Lambda_*^2/m$.\footnote{For a detailed discussion of the Efimov effect
for finite scattering length and applications to ultracold
atoms, see Ref.~\cite{Braaten:2004rn}.}.
The discrete scaling symmetry becomes also manifest in the log-periodic
dependence of three-body observables on the scattering length.
This log-periodic behavior is the hallmark signature of a
renormalization group limit cycle. 
It has been observed experimentally 
in the three-body recombination spectra of ultracold atomic gases 
close to a Feshbach resonance~\cite{Ferlaino:2010viw,PhysRevA.93.022707}.

\begin{figure}[t]
\begin{center}
\includegraphics[width=8cm,clip=]{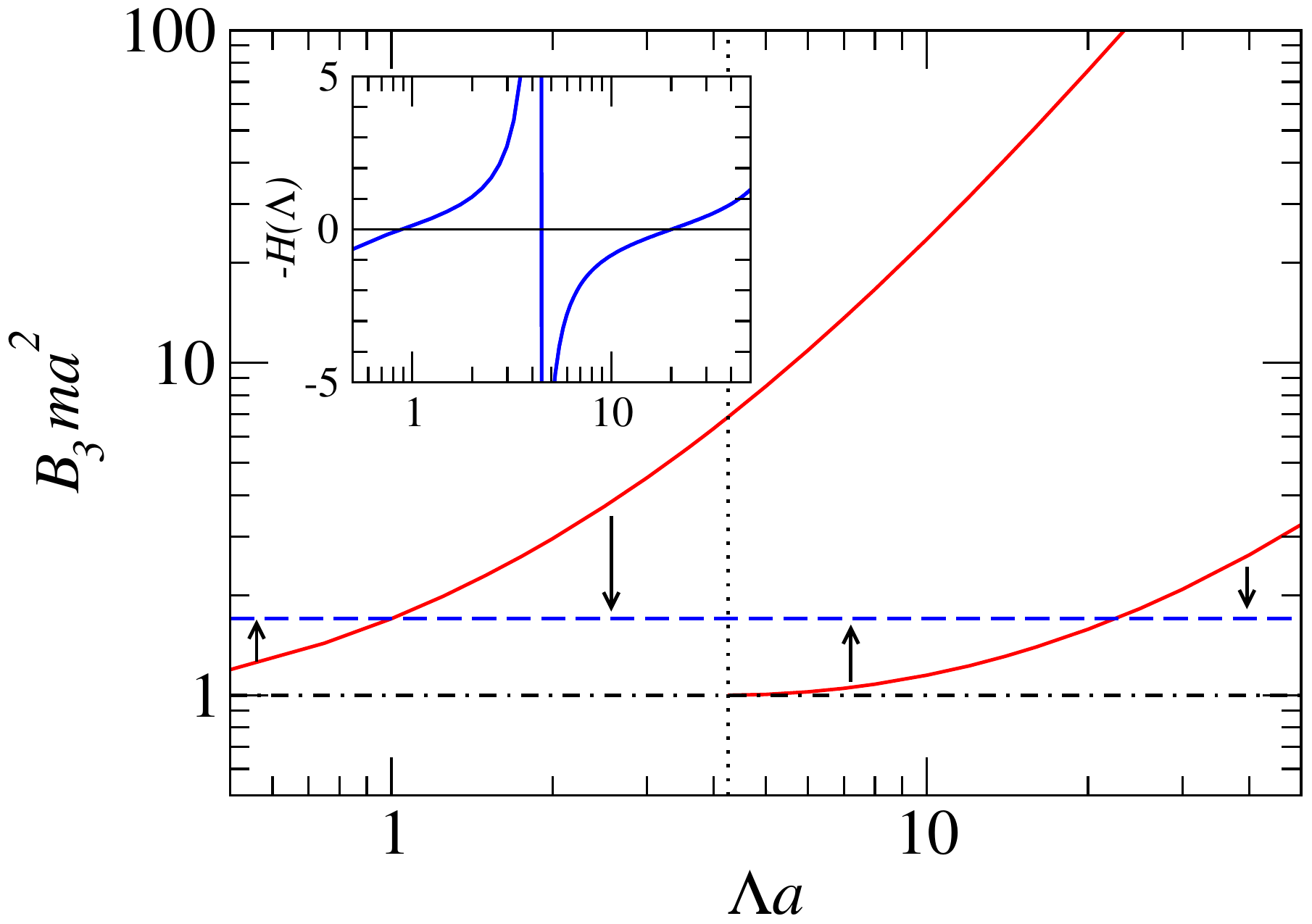}
\end{center}
\caption{Unrenormalized three-body energies $B_3$ as a function of
the momentum cutoff $\Lambda$ (solid lines).  
The dotted line indicates the cutoff where a new three-body state appears 
at the particle-dimer threshold (dash-dotted line). The dashed line 
shows a hypothetical renormalized energy. The inset shows the running
of the three-body force $d_0(\Lambda) \sim - H(\Lambda)$ with $\Lambda$.}
\label{fig:B3lambda}
\end{figure}
The physics of the renormalization procedure is illustrated in 
Fig.~\ref{fig:B3lambda} where we show the unrenormalized 
three-body binding energies $B_3$ in the case of positive scattering length
as a function of the cutoff $\Lambda$ (solid line).
As the cutoff $\Lambda$ is increased, $B_3$ increases
asymptotically as $\Lambda^2$. At a certain cutoff 
(indicated by the dotted line), a new bound state appears at the 
boson-dimer threshold. This pattern repeats every time the cutoff
increases by the discrete scaling factor $\exp(\pi/s_0)$. 
Now assume that we adopt 
the renormalization condition that the shallowest state should have a constant 
energy given by the dashed line. At small values of the cutoff,
we need an attractive three-body force to increase the binding energy 
of the shallowest state as indicated by the arrow. As the cutoff is increased
further, the required attractive contribution becomes smaller and around
$\Lambda a =1.1$ a repulsive three-body force is required (downward arrow). 
Around $\Lambda a=4.25$, a new three-body state appears at threshold
and we cannot satisfy the renormalization condition by keeping the first 
state at the required energy anymore. The number of bound states has changed
and there is a new shallow state in the system. At this
point the three-body force 
turns from repulsive to attractive to move the
new state to the required energy.  The corresponding running of the 
three-body force with the cutoff $\Lambda$ is shown in the inset.
After renormalization, the first state is still present as a deep state 
with large binding energy, but for
threshold physics its presence can be ignored. This pattern goes on 
further and further as the cutoff is increased~\cite{Bedaque:1998km}.

\section{Effective field theory for nuclear few-body systems}
\label{sec:EFT-Nucleons}

\subsection{Overview}

Depending on the physics one wishes to describe, there are several effective 
field theories for low-energy nuclear physics to choose from.  They differ in 
the set of effective degrees of freedom, their expansion point (typical 
low-energy scale) and range of applicability.  Chiral effective field theory 
includes nucleons and pions and is designed as an expansion about the so-called 
``chiral limit,'' \ie, the scenario where the quark masses are exactly zero 
such that the pions emerge as exactly massless Goldstone bosons from the 
spontaneous breaking of chiral symmetry.  In reality, the quark-masses are 
nonzero such that the pions become ``pseudo-Goldstone'' bosons with a small 
(compared to typical QCD scales like $\mrho$ or $\MN$) mass $\mpi$.  Chiral EFT 
takes this as a typical low scale so that its power counting is designed for 
momenta of the order $Q\sim\mpi$; we come back to this in 
Sec.~\ref{sec:EFT-Chiral}.

For momenta much smaller than $\mpi$, explicit pion 
exchange cannot be resolved such that these can be regarded as integrated out, 
much like we did explicitly for the pseudoscalar toy model in 
Sec.~\ref{sec:EFT-IntOut-2}.  The resulting ``pionless'' theory has, up to 
long-range forces that we consider in Sec.~\ref{sec:EFT-EM}) only contact 
interactions between nucleons left.  These contact interactions parameterize not 
only unresolved pion exchange, but also that of heavier mesons, for which 
contact terms already exist in Chiral EFT.  Pionless EFT is formally very 
similar to the few-boson EFT discussed in Sec.~\ref{sec:EFT-Bosons}, and it 
despite its simplicity it gives rise to surprisingly rich physics, as we will 
show in this section.

\subsection{Pionless effective field theory}

The neutron-proton S-wave scattering lengths are experimentally determined to 
be about $5.4~\fm$ in the \ThreeSOne channel, and ${-}23.7~\fm$ in the \OneSNot 
channel.  What is special about these numbers is that they are large compared 
to the typical nuclear length scale determined by the pion Compton wavelength, 
$\mpi^{-1}\sim1.4~\fm$.  This estimate comes from the long-range component of 
the nuclear interaction being determined by one-pion exchange.  Na\"ively, if 
we consider the low-energy limit ($NN$ center-of-mass momentum going to zero) 
we expect that we can integrate out the pions and end up with a contact 
interaction scaling with the inverse pion mass, and thus a perturbative EFT 
reproducing natural-sized scattering lengths.  The fact that this is not the 
case is typically interpreted as nature ``choosing'' the fine-tuned scenario 
outlined in Sec.~\ref{sec:EFT-FineTuning}.  In this case, pion 
exchange\footnote{As discussed in Sec.~\ref{sec:EFT-Chiral} this actually has to 
be the exchange of two or more pions, as one-pion exchange does not contribute 
to S-wave scattering at zero energy.} combines with shorter range interactions 
to yield the large S-wave scattering lengths (and the deuteron as an 
unnaturally shallow bound state), implying that nuclear physics is a strongly 
coupled and thus nonperturbative system at low energies.  This is what allows 
us to write down an EFT that is closely related to the one that describes 
strongly interacting bosons.  What governs the physics of low-energy 
observables is to a good appropriation just the fact that the scattering 
lengths are large, so we end up with a short-range EFT much like the one for 
bosons encountered in Sec.~\ref{sec:EFT-Bosons}.  Some rather technical new 
features arise from the fact that nucleons are fermions with spin and isospin.

\subsection{The two-nucleon S-wave system}

The leading-order Lagrangian for pionless EFT can be written as

\begin{equation}
 \LL
 = N^\dagger \left(\ii\partial_0 + \frac{\Laplace}{2\MN} + \cdots\right)N \\
 - C_{0,s} (N^T \hat{P}_s N)^\dagger(N^T \hat{P}_s N) \\
 - C_{0,t} (N^T \hat{P}_t N)^\dagger(N^T \hat{P}_t N)
 + \cdots \,,
\label{eq:L-NN}
\end{equation}
with projectors
\begin{equation}
 (\hat{P}_t)^i = \sigma^2\sigma^i\tau^2 / \sqrt8 \mathtext{,}
 (\hat{P}_s)^\lambda = \sigma^2\tau^2\tau^\lambda/\sqrt8
\label{eq:P-t-s}
\end{equation}
such that $C_{0,s}$ and $C_{0,t}$ refer to the \OneSNot and \ThreeSOne $NN$ 
channels, respectively.  As in previous sections of this chapter, we use 
$\sigma^i$ to denote the Pauli matrices in spin space, and write 
$\idxx{\sigma^i}\alpha\beta$ to refer to their individual entries (with the
upper index referring to the row).  Conversely, we use the notation 
$\idxx{\tau^\lambda}ab$ in isospin space.

With the usual cartesian indices $i,\lambda = 1,2,3$ the projectors for given 
$i$ or $\lambda$ give somewhat unusual combinations of individual states.  For 
example, $np$ configurations are completely 
contained in the $\lambda=3$ isospin component, whereas $nn$ and $pp$ are 
obtained from linear combinations $1\pm\ii2$,  In other words, in order to 
separate the physical states (and likewise to get spin-1 states with $m=0,\pm1$ 
quantum numbers) one should work instead with a spherical basis.  For example, 
if one wants to include isospin-breaking terms, it is convenient to work with 
the projectors
\begin{equation}
 (\tilde{P}_s)^{-1}
 = \frac{1}{\sqrt{2}}\left[(\hat{P}_s)^1-\ii (\hat{P}_s)^2\right]
 \mathtext{,}
 (\tilde{P}_s)^{0}
 = (\hat{P}_s)^3
 \mathtext{,}
 (\tilde{P}_s)^{+1}
 = {-}\frac{1}{\sqrt{2}}\left[(\hat{P}_t)^1+\ii (\hat{P}_s)^2\right] \,.
\label{eq:P-s-spherical}
\end{equation}
Otherwise, since the difference is a unitary rotation, the choice of basis is 
arbitrary.

\subsubsection{Spin and isospin projection}

To understand why the projectors have been defined as in Eq.~\eqref{eq:P-t-s}, 
it is instructive to calculate the tree-level contribution to the amplitude in a 
given channel.  With all spin and isospin vertices written out, the Feynman rule 
for the four-nucleon vertex in the \ThreeSOne channel is
\begin{equation}
 \parbox{7.5em}{\centering\includegraphics[width=6.5em]{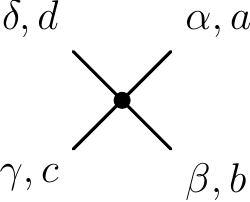}}
 \sim \; \ii \frac{C_{0,t}}{8}
 \idxx{\sigma^i\sigma^2}\alpha\beta
 \idxx{\tau^2}ab
 \idxx{\sigma^2\sigma^i}\gamma\delta
 \idxx{\tau^2}cd \,,
\label{eq:C0-Vertex}
\end{equation}
which is obtained by simply writing out the $(\hat{P}_t)^i$ from 
Eq.~\eqref{eq:P-t-s}.  Furthermore, this diagram has an associated combinatorial 
factor $4$ because there are two possibilities each to contract the in- and 
outgoing legs with external nucleon fields.

In order to calculate the T-matrix, we have to write out the Lippmann--Schwinger 
equation with all indices (and symmetry factors) and then apply the appropriate 
projectors.  For ${}^3S_1$ and isospin $0$, the result should have two free 
spin-$1$ indices, which we label $k$ and $j$ for the in- and outgoing side, 
respectively.  The inhomogeneous term is just the vertex~\eqref{eq:C0-Vertex} 
with an additional factor $4$.  Applying the projectors, we get
\begin{multline}
 \frac{1}{\sqrt8} \idxx{\sigma^2\sigma^j}\beta\alpha \idxx{\tau^2}ba
 \times 4
 \times \ii\frac{C_{0,t}}{8}\idxx{\sigma^i\sigma^2}\alpha\beta \idxx{\tau^2}ab
 \idxx{\sigma^2\sigma^i}\gamma\delta \idxx{\tau^2}cd
 \times \frac{1}{\sqrt8} \idxx{\tau^2}dc
 \idxx{\sigma^k\sigma^2}\delta\gamma \\
 = \ii\frac{C_{0,t}}{16} \, \Tr\big(\sigma^j\sigma^i\big)
 \, \Tr\big(\tau^2\tau^2\big)
 \, \Tr\big(\sigma^i\sigma^k\big) \, \Tr\big(\tau^2\tau^2\big)
 = \ii C_{0,t} \, \delta^{jk} \,,
\end{multline}
where we have used the well-known of the Pauli matrices.  This is 
exactly the expected result: the projectors~\eqref{eq:P-t-s} have been 
constructed to give this.  The projection of other, more complicated diagram 
works in the same way.  Albeit somewhat tedious, it is technically 
straightforward.  For higher partial waves, one of course has to take into 
account the coupling of spin and orbital angular momentum.

\subsubsection{Dibaryon fields}
\label{sec:EFT-Dibaryons}

Just like for bosons, it is convenient to introduce auxiliary dimer---now 
called dibaryon---field for each of the $NN$ S-wave states.  This is done by 
rewriting the Lagrangian~\eqref{eq:L-NN} as
\begin{multline}
 \LL
 = N^\dagger \left(\ii\partial_0 + \frac{\Laplace}{2\MN} + \cdots\right)
 - t^{i\dagger}\left[\sigt
 + \left(\ii \partial_0+\frac{\vNabla^2}{4\MN}\right)\right]t^i
 + \yt\left[t^{i\dagger}\left(N^T P^i_t N\right)+\hc\right] \\
 - s^{\lambda\dagger}\left[\sigs +
 \left(\ii \partial_0+\frac{\vNabla^2}{4\MN}\right)\right]s^\lambda
 + \ys\left[s^{\lambda\dagger}\left(N^T P^\lambda_s N\right)+\hc\right]
 + \cdots \,,
\label{eq:L-NN-d}
\end{multline}
where $t$ ($s$) denotes a \ThreeSOne (\OneSNot) dibaryon field and the 
projection operators $P_{s,t}$ are as defined in Eq.~\eqref{eq:P-t-s}.  The 
``bare'' dibaryon propagators are just $\ii/\sigst$, while the full 
leading-order expressions are obtained by summing up all nucleon bubble 
insertions.  This resummation, which without dibaryon fields gives the $NN$ 
T-matrix as a bubble chain, reflects the fact we need to generate shallow 
states (the bound deuteron and the virtual \OneSNot state) to account for the 
unnaturally large $NN$ scattering lengths.  Pionless EFT is designed to capture 
this feature.

Omitting spin-isospin factors for simplicity, the resummed propagators are
\begin{equation}
 \ii\Delta_{\st}(p_0,{\mathbf{p}})
 = \frac{-\ii}{\sigst + \yst^2 I_0(p_0,{\mathbf{p}})} \,,
\end{equation}
where
\begin{multline}
 I_0(p_0,{\mathbf{p}}) = \MN\int^\Lambda \frac{\dd^3q}{(2\pi)^3}
 \frac{1}{\MN p_0 - {\mathbf{p}}^2/4 - \vq^2 + \ii\eps}
 = {-}\frac{\MN}{4\pi}
 \left(\frac{2\Lambda}{\pi} - \sqrt{\frac{{\mathbf{p}}^2}{4}-\MN p_0-\ii\eps}\right)
 +\OO(1/\Lambda)
\label{eq:I0-cutoff}
\end{multline}
is the familiar bubble integral regularized with a momentum cutoff.  The cutoff 
dependence is absorbed into the parameters $\yst$ and $\sigst$ to obtain the 
renormalized propagators.  Attaching external nucleon legs gives the $NN$ 
T-matrix,
\begin{equation}
 \ii T_\st(k)
 = (\ii \yst)^2\, \ii\Delta_\st\!\left(p_0=k^2/\MN,{\mathbf{p}}=\vZero\right)
 = \frac{4\pi}{M_N}\frac{\ii}{k\cot\delta_\st(k) - \ii k}\,,
\label{eq:Tnd}
\end{equation}
so we can match to the effective range expansions for $k\delta_\st(k)$.  At 
leading order, the renormalization condition is to reproduce
$k\cot\delta_\st(k) = {-}1/{a_\st} + \OO(k^2)$, which gives
\begin{equation}
 \frac{4\pi\sigst}{\MN\yst^2} = {-}\frac{1}{a_\st} + \frac{2\Lambda}{\pi} \,.
\label{eq:NN-renorm-0}
\end{equation}
Instead of this standard choice of the expansion around the zero-energy 
threshold, it is convenient to expand the \ThreeSOne channel around the 
deuteron pole.\footnote{The shallow deuteron bound-state pole is within the 
radius of convergence of the effective range expansion. The deuteron binding 
momentum is $\gamt=1\at + \cdots$, where the ellipses include corrections from 
the effective range (and higher-order shape parameters).}  This is
\begin{equation}
 k\cot\delta_t(k) = \gamt + \dfrac{\rhot}2\big(k^2+\gamt^2\big) + \cdots \,,
\end{equation}
where $\gamt = \sqrt{\mathstrut\MN \Bd}\simeq 45.7$ is the deuteron binding 
momentum and $\rhot\simeq 1.765$ is the ``deuteron effective range.''  This 
choice, which sets
\begin{equation}
 \frac{4\pi\sigt}{\MN\yt^2} = {-}\gamt + \frac{2\Lambda}{\pi} \,,
\label{eq:NN-renorm-deut}
\end{equation}
gets the exact deuteron pole position right at leading order, but is equivalent 
to the choice in Eq.~\eqref{eq:NN-renorm-0} up to range corrections.

\paragraph{Wavefunction renormalization}

The residue at the deuteron pole gives the deuteron wavefunction 
renormalization.  We find
\begin{equation}
 Z_t^{-1} = \ii\frac{\partial}{\partial p_0}
 \left.\frac{1}{\ii\Delta_t(p_0,{\mathbf{p}})}\right|_{p_0=-\frac{\gamt^2}{\MN},\,{\mathbf{p}}=0}
 = \frac{\MN^2 \yt^2}{8\pi\gamt}
 \;\Rightarrow\;
 \sqrt{Z_t} = \frac{1}{\yt} \frac{\sqrt{8\pi\gamt\mathstrut}}{\MN}
\label{eq:Zd}
\end{equation}
for the renormalization as in Eq.~\eqref{eq:NN-renorm-deut}.  If we 
directly consider the (off-shell) T-matrix near the pole, we find
\begin{spliteq}
 T(E) &= \frac{4\pi}{\MN}
 \frac{1}{-\gamt+\sqrt{-\MN E-\ii\eps\mathstrut}}
 = {-}\frac{4\pi}{\MN}
 \frac{\sqrt{-\MN E-\ii\eps\mathstrut}+\gamt}{-\MN E-\gamd^2}
 \mathtext{for $\eps\to0$} \\
 &\sim
 {-}\frac{8\pi\gamt}{\MN^2} \frac{1}{E+\dfrac{\gamt^2}\MN}
 \mathtext{as $E\to-\gamt^2/\MN$} \,.
\label{eq:T-deut-pole}
\end{spliteq}
Comparing to the standard factorization at the pole,\footnote{The minus sign in 
Eq.~\eqref{eq:T-factorization} is a consequence of the convention we use here 
for the T-matrix.}
\begin{equation}
 T(k,p;E) = {-}\frac{B^\dagger(k)B(p)}{E+E_B}
 + \text{terms regular at $E=-E_B$} \,,
\label{eq:T-factorization}
\end{equation}
we can read off from Eq.~\eqref{eq:T-deut-pole} that
\begin{equation}
 B(p) = \sqrt{\frac{8\pi\gamt}{\MN^2}} = \yt\sqrt{Z_t} \,,
\label{eq:B-deut}
\end{equation}
independent of momentum at this order.

\subsection{Three nucleons: scattering and bound states}

As done in Sec.~\ref{sec:EFT-ThreeBosons} for bosons, the dimer/dibaryon 
formalism allows for a particularly intuitive and simple description of the 
three-body system.  Looking at nucleon-deuteron S-wave scattering\footnote{We 
work in the isospin-symmetric theory here, but in the absence of 
electromagnetic interactions (discussed in Sec.~\ref{sec:EFT-EM}), the nucleon 
here should be thought of as a neutron.}, we find that the spin $1$ of the 
deuteron can couple with the spin $1/2$ of the nucleon to a total spin of 
either $3/2$ or $1/2$.  These two cases are referred to as the quartet and 
doublet channel, respectively.

\subsubsection{Quartet channel}
\label{sec:EFT-NdQuartet}

\begin{figure}[htbp]
\centering
\includegraphics[clip,width=0.5\textwidth]{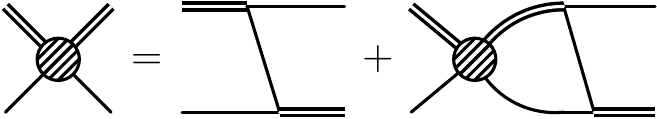}
\caption{$Nd$ quartet-channel integral equation.  Nucleons and deuterons are 
represented as single and double lines, respectively.  The blob represents the 
T-matrix.}
\label{fig:nd-IntEq-Q}
\end{figure}

In the quartet-channel, the spins of all three nucleons have to be aligned to 
produce the total spin $3/2$.  This means that only the deuteron field can 
appear in intermediate states, and the Pauli principle excludes a three-body 
contact interaction without derivatives.  The resulting integral equation for 
the $Nd$ T-matrix is shown diagrammatically in Fig.~\ref{fig:nd-IntEq-Q}.  
Compared to Fig.~\ref{fig:inteq12}, we use a different convention here were the 
T-matrix blob is drawn to the left of the nucleon exchange, and we denote in- 
and outgoing momenta (in the $Nd$ center-of-mass frame) by $\pm\vk$ and 
$\pm{\mathbf{p}}$, respectively.\footnote{Unlike what is done in 
Sec.~\ref{sec:EFT-Bosons}, we also read diagrams from left (incoming particles) 
to right (outgoing particles).  Both conventions can be found in the 
literature.}  The energy and momentum dependence are exactly the same as for 
bosons, but we have to include the additional spin-isospin structure from the 
vertices.  Doing this, the result in its full glory reads:
\begin{multline}
 \idxx{\ii T_q^{ij}}{\beta b}{\alpha a}(\vk,{\mathbf{p}};E)
 = -\frac{\ii\MN\yt^2}{2}\,\idxx{\sigma^j\sigma^i}\beta\alpha\delta^b_a
 \,\frac{1}{\vk^2+\vk\cdot{\mathbf{p}}+{\mathbf{p}}^2-\MN E-\ii\eps}\\
 +\int\frac{\dd^3 q}{(2\pi)^3} \,
 \Delta_t\left(E-\frac{\vq^2}{2\MN},\vq\right)
 \idxx{\ii T_q^{ik}}{\gamma c}{\alpha a}(E;\vk,\vq)\\
 \times\frac{\MN\yt^2}{2}
 \frac{\idxx{\sigma^j\sigma^k}\beta\gamma\delta^b_c}
 { \vq^2+\vq\cdot{\mathbf{p}}+{\mathbf{p}}^2-\MN E-\ii\eps} \,.
\label{eq:nd-IntEq-Q-raw}
\end{multline}
This unprojected amplitude carries spin and isospin indices for the various 
fields in the initial and final states.  To select the overall spin $3/2$ 
contribution, we take linear combinations as in Eq.~\eqref{eq:P-s-spherical} to 
select the maximal projections for the in- and outgoing deuterons and 
$\alpha=\beta=1$ to get nucleons with spin orientation ${+}1/2$.  We also set 
$a=b=2$ to select neutrons.  Altogether, this gives
\begin{equation}
 \idxx{\sigma^j\sigma^i}\beta\alpha\delta^b_a \rightarrow 2 \,,
\end{equation}
in the inhomogeneous term, and the same factor for the integral part.  The 
fully projected quartet-channel amplitude is
\begin{equation}
 T_q = \frac{1}{2}\idxxx{T_q^{11} + \ii\left(T_q^{12} - T_q^{21}\right)
 + T_q^{22}}{12}{12} \,.
\label{eq:T-q}
\end{equation}

\begin{prob}
\emph{Exercise:} Work out the details leading to Eq.~\eqref{eq:T-q}.
\end{prob}

Finally, the S-wave projection of Eq.~\eqref{eq:nd-IntEq-Q-raw} is done by 
applying the operator $\frac12\int_{-1}^{1}\dd\cos\theta$, where $\theta$ is 
the angle between $\vk$ and ${\mathbf{p}}$.  Introducing a momentum cutoff $\Lambda$, 
the resulting equation can be solved numerically by discretizing the 
remaining one-dimensional integral.  From the result, which we denote by 
$T_q^0$, we can calculate observables like scattering phase shifts,
\begin{equation}
 \delta_{q}(k) = \frac{1}{2\ii}
 \ln\!\left(1+\frac{2\ii k\MN}{3\pi} Z_t T_q^0(k,k;E_k)\right)
 \mathtext{,} E_k = \frac{3k^2}{4\MN}-\frac{\gamt^2}{\MN} \,,
\label{eq:delta-q}
\end{equation}
or the $Nd$ scattering length:
\begin{equation}
 a_q = \frac{\MN}{3\pi}\lim\nolimits_{k\to0} Z_t T_q^0(k,k;E_k) \,.
\label{eq:and-2}
\end{equation}
Note that we have not absorbed the wavefunction renormalization $Z_t$ into 
$T_q^0$ but instead chose to keep it explicit in Eqs.~\eqref{eq:delta-q} 
and~\eqref{eq:and-2}

\subsubsection{Doublet channel}
\label{sec:EFT-NdDoublet}

\begin{figure}[htbp]
\centering
\includegraphics[clip,width=0.65\textwidth]{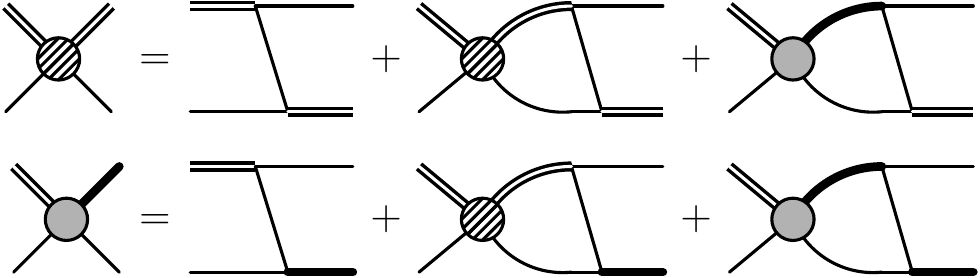}
\caption{$Nd$ doublet-channel integral equation.  As in 
Fig.~\ref{fig:nd-IntEq-Q}, nucleons and deuterons are drawn as single and 
double lines, respectively.  Additionally, we represent the \OneSNot dibaryon 
as a thick line.  The hatched and shaded blobs are the two components of the 
doublet-channel $Nd \to Nd$ T-matrix.}
\label{fig:nd-IntEq-D}
\end{figure}

The doublet channel (total spin $1/2$) has a richer structure, since now also 
the \OneSNot dibaryon can appear as an intermediate state.  The result is a 
coupled channel integral equation, shown diagrammatically in 
Fig.~\ref{fig:nd-IntEq-D}.  We skip here the technicalities of the spin-isospin 
projection (for details, see Ref.~\cite{Konig:2014ufa} and earlier references 
therein) and merely quote the result in a compact notation:
\begin{equation}
 \twodvec{T_{d,1}^0 \\ T_{d,2}^0}
 = \twodvec{\MN\yt^2/2 \\ {-}3\MN\yt\ys/2} K
 + \twodmat{D_t & 0 \\ 0 & D_s}
 \twodmat{{-}\MN\yt^2/2 & 3\MN\yt\ys/2 \\ 3\MN\ys\yt/2 & {-}\MN\ys^2/2} K
 \otimes \twodvec{T_{d,1}^0 \\ T_{d,2}^0} \,,
\label{eq:nd-IntEq-noH}
\end{equation}
where $K$ is the ``kernel'' function 
\begin{equation}
 K(k,p;E) = \frac{1}{2kp}\ln\!\left(
 \frac{k^2+p^2+kp-\MN E-\ii\eps}{k^2+p^2-kp-\MN E-\ii\eps}\right) \,,
\label{eq:KS}
\end{equation}
\begin{equation}EFT-
 D_\st(q;E) = \Delta_\st\!\left(E-\frac{q^2}{2\MN};q\right) \,,
\label{eq:D-st}
\end{equation}
and the integral operation
\begin{equation}
 A \otimes B \equiv \frac1{2\pi^2}
 \int_0^\Lambda\dd q\,q^2\,A(\ldots,q)B(q,\ldots)
\label{eq:SH-int}
\end{equation}
has to be applied within each block.  Just like the quartet-channel equation, 
Eq.~\eqref{eq:nd-IntEq-noH} can be solved numerically by discretizing the 
integrals, with the additional complication that we now have a $2\times2$ block 
matrix.  The T-matrix likewise becomes a $2$-block vector, the upper part 
of which is gives the physical $Nd \to Nd$ amplitude.\footnote{Note that this 
vector is one part of the more general full off-shell amplitude, which is 
a $2\times2$ block matrix including the two combinations of dibaryon legs that 
do not appear in Fig.~\ref{fig:nd-IntEq-D}.}

\begin{prob}
\emph{Exercise:} Express the fully projected integral equation for the quartet 
channel amplitude $T_q^0$ using the compact notation based on 
Eqs.~\eqref{eq:KS}, \eqref{eq:D-st}, and~\eqref{eq:SH-int}.
\end{prob}

\paragraph{Leading-order three-nucleon force}

Studying the doublet-channel solution as a function of increasing UV cutoff 
$\Lambda$, one finds that there is no stable limit as $\Lambda\to\infty$.  
Instead, the amplitude changes wildly as $\Lambda$ is varied.  This is much 
unlike the quartet-channel case, which shows a rapid convergence with $\Lambda$.

The origin of this behavior was explained by 
Bedaque~\etal~\cite{Bedaque:1999ve}.  The behavior for large $\Lambda$ is 
governed by large momenta, which means that infrared scales like the scattering 
lengths do not matter.  Indeed, one finds that $D_t(E;q)$ and $D_s(E;q)$ have 
the same leading behavior as $q\to\infty$.  To analyze the asymptotic behavior 
of the amplitude we can thus go to the $SU(4)$ spin-isospin symmetric limit and 
set $D_t = D_t \equiv D$ as well as $\yt = \ys \equiv \y$.  In this limit, the 
two integral equations in Eq.~\eqref{eq:nd-IntEq-noH} can be decoupled by 
defining $T_{d,\pm} = T_{d,1} \pm T_{d,2}$.  For $T_{d,+}$ we find the integral 
equation
\begin{equation}
 T_{d,+}^0(k,p;E) = {-}\MN\y^2 K(k,p;E)
 + \MN\y^2 \int_0^\Lambda \frac{\dd q}{2\pi^2}
 \, q^2 \, K(k,q;E) D(q;E) T_{d,+}^0(q,p;E) \,,
\end{equation}
which is formally exactly the same as the three-boson integral equation, 
Eq.~\eqref{BHvK}, in the absence of a three-body force.  As discussed in 
Sec.~\ref{sec:EFT-ThreeBosons}, this equation does not have a unique solution 
in the limit $\Lambda\to\infty$.  Since $T_{d,1/2}$ are linear combinations of 
involving $T_{d,+}$, they inherit the same behavior.  But the cure is now 
obvious: a three-nucleon force, which by naïve counting would only enter at 
higher orders, has to be promoted to leading order in order to make 
Eq.~\eqref{eq:nd-IntEq-noH} well defined.  This three-nucleon force, like the 
asymptotic amplitudes, is $SU(4)$ symmetric (invariant under arbitrary 
spin-isospin rotations) and can be written as
\begin{equation}
 \mathcal{L}_3 = \frac{h_0}{3}N^\dagger\left[\yt^2\,
 t^{i\dagger} t^j \sigma^i \sigma^j+\ys^2\,s^{A\dagger} s^B \tau^A\tau^B
 - \yt\ys\left(t^{i\dagger} s^A \sigma^i \tau^A + \hc\right) \right]N \,,
\label{eq:L-3}
\end{equation}
where the cutoff-running of the three-nucleon coupling is analogous to 
what we derived for bosons:
\begin{equation}
 h_0 = \frac{\MN H(\Lambda)}{\Lambda^2} \,.
\end{equation}
Including this, the coupled doublet-channel integral equation becomes
\begin{equation}
 \twodvec{T_{d,1}^0 \\ T_{d,2}^0}
 = \twodvec{\MN\yt^2/2 \\ {-}3\MN\yt\ys/2} K
 + \twodmat{D_t & 0 \\ 0 & D_s}
 \twodmat{{-}\MN\yt^2/2(K + h_0) & 3\MN\yt\ys/2 (K + h_0) \\
 3\MN\ys\yt/2 (K + h_0) & {-}\MN\ys^2/2 (K + h_0)}
 \otimes \twodvec{T_{d,1}^0 \\ T_{d,2}^0} \,.
\label{eq:nd-IntEq}
\end{equation}
We stress that the requirement to include this three-body force at 
leading-order is a feature of the nonperturbative physics that can be traced 
back to the large $NN$ scattering lengths.  All loop diagrams obtained by 
iterating the integral equation are individually finite as $\Lambda\to\infty$, 
yet their infinite sum does not exist in that limit unless the $h_0$ contact 
interaction is included.

\paragraph{The Phillips line}

\begin{figure}[tb]
\centering
\includegraphics[width=0.65\textwidth,clip]{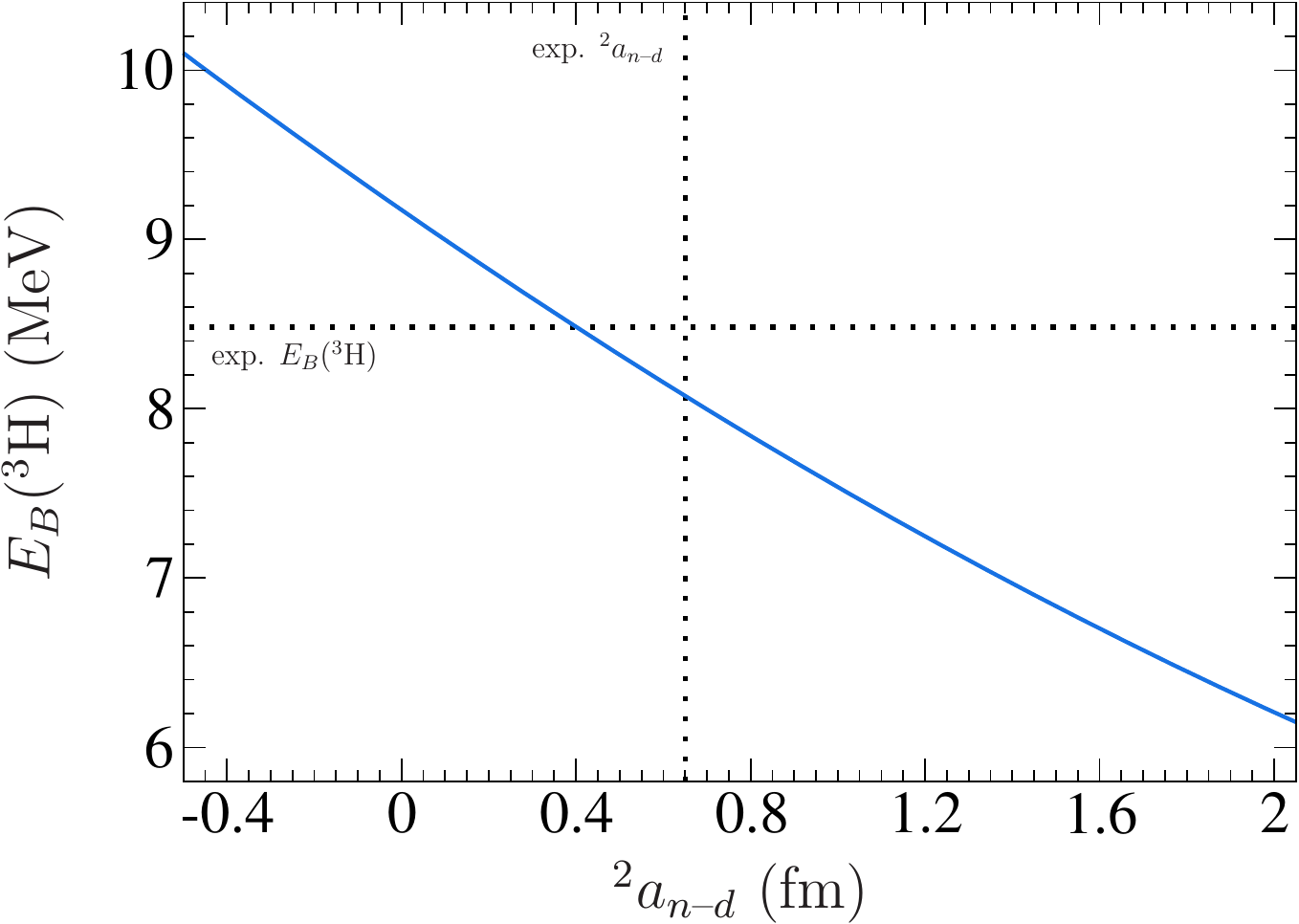}
\caption{Correlation (Phillips line) between \Triton binding energy (in \MeV)
and doublet $nd$ scattering length (in \fm) in leading-order pionless EFT.}
\label{fig:Phillips}
\end{figure}

The form of $H(\Lambda)$ is given by Eq.~\eqref{H-Lambda}, but since there is a 
prefactor that depends on the details of the regularization scheme, in practice 
one has to determine the appropriate value numerically after choosing a cutoff 
$\Lambda$.  This requires a three-body datum as input, which is conveniently 
chosen to be the triton binding energy---the \ThreeH bound state corresponds to 
a pole in $T_d$ at $E = {-}E_B(\ThreeH)$, \cf~Sec.~\ref{sec:EFT-3BObs}---or the 
doublet-channel $nd$ scattering length (or, in principle, a phase shift at some 
fixed energy).  Once one of these is fixed, the rest can be predicted.  In 
particular, this means that the existence of a single three-body parameter in 
pionless EFT at leading order provides a natural explanation of the Phillips 
line, \ie, the observation that various phenomenological potentials, which 
were all tuned to produce the same $NN$ phase shifts, gave different results 
for the triton binding energy and doublet S-wave scattering length, which 
however are strongly correlated.  In our framework, we can obtain the line 
shown in Fig.~\ref{fig:Phillips} by fitting $H(\Lambda)$ to different values of 
the scattering length and then calculating $E_B(\ThreeH)$ (or vice versa).  
Alternatively, one can find the same curve by setting $h_0$ to zero and varying 
$\Lambda$ to move along the line.

\medskip
With this, we conclude our discussion the three-nucleon system.  For more 
details, like calculations beyond leading order, we encourage the reader to 
encourage the literature.

\section{Beyond short-range interactions: adding photons and pions}
\label{sec:EFT-Beyond}

\subsection{Electromagnetic interactions}
\label{sec:EFT-EM}

So far, we have studied only effective theories where the particles (atoms of 
nucleons) interact solely via short-range force (regulated contact 
interactions).  While this is, as we have argued, sufficient to describe the 
strong nuclear interactions at (sufficiently low) energies where the EFT is 
valid, the electromagnetic interaction does not fit in this scheme.  
Nevertheless, since almost all systems of interest in low-energy nuclear 
physics involve more than one proton, the inclusion of such effects 
is of course important.

Any coupling of photons to charged particles has to be written down in a 
gauge-invariant way.  A natural way to ensure gauge invariance is to replace 
all derivatives in the effective Lagrangian with covariant ones, \ie,
\begin{equation}
 \partial_\mu \rightarrow D_\mu = \partial_\mu + \ii eA_\mu \hat{Q} \,,
\label{eq:D-mu}
\end{equation}
where $\hat{Q}$ is the charge operator (for nucleons, for example, 
$\hat{Q}_N=(\one+\tau_3)/2$).  Moreover $e^2 = 4\pi\alpha$ defines the electric 
unit charge in terms of the fine structure constant $\alpha\approx1/137$.

Since in the EFT there is an infinite tower of contact interactions with an 
increasing number of derivatives, we also get an infinite number of photon 
vertices.  Still, merely plugging in the covariant derivative is not enough 
(after all, this is called \emph{minimal} substitution); it is possible to write 
down terms which are gauge invariant by themselves, and for the EFT to be 
complete these have to be included as well.  Before we come back to this in 
Sec.~\ref{sec:EFT-ExtCurrents}, let us first see what we get from gauging the 
derivatives in the nucleon kinetic term:
\begin{equation}
 N^\dagger \left(\ii\partial_t + \frac{\Laplace}{2\MN} + \cdots\right)N
 \rightarrow
 N^\dagger \left(\ii D_t + \frac{\vD^2}{2\MN} + \cdots\right)N \,.
\end{equation}
In addition to this, we also have to include the photon 
kinetic and gauge fixing terms to complete the electromagnetic sector.  A 
convenient choice for the our nonrelativistic framework is Coulomb gauge, \ie, 
demanding that $\vNabla\cdot\vA=0$.  A covariant way to write is condition is
\begin{equation}
 \partial_\mu A^\mu-\eta_\mu\eta_\nu\partial^\nu A^\mu = 0
\label{eq:Coulomb-Gauge}
\end{equation}
with the timelike unit vector $\eta^\mu=(1,0,0,0)$.  Hence, we add to our 
effective Lagrangian the term
\begin{equation}
 \mathcal{L}_\mathrm{photon} = -\frac14 F_{\mu\nu}F^{\mu\nu} -\frac{1}{2\xi}
 \left(\partial_\mu A^\mu-\eta_\mu\eta_\nu\partial^\nu A^\mu\right)^2 \,.
\label{eq:L-photon}
\end{equation}

\subsubsection{The Coulomb force}

From Eq.~\eqref{eq:L-photon} we get the equation of motion for the photon field 
as
\begin{equation}
 \left[\dAlem g^{\mu\nu} - \partial^\mu\partial^\nu
 + \frac{1}{\xi}\left(\partial^\mu\partial^\nu
 - \eta^\nu\eta_\kappa\partial^\mu\partial^\kappa 
 - \eta^\mu\eta_\lambda\partial^\lambda\partial^\nu
 + \eta^\mu\eta^\nu\eta_\lambda\eta_\kappa\partial^\lambda\partial^\kappa
 \right)\right] A_\nu(x) = 0 \,.
\end{equation}
The photon propagator is defined as the Feynman Green's function for the 
differential operator acting on $A_\nu(x)$.  Writing down the general solution 
in momentum space and choosing $\xi = 0$ at the end (recall that it is an 
artificial parameter introduced through enforcing the gauge condition in the  
path integral~\cite{Peskin:1995}), we get
\begin{equation}
 D_\gamma^{\mu\nu}(k) = \frac{-\ii}{k^2+\ii\eps}
 \left(g^{\mu\nu}+\frac{k^\mu k^\nu + k^2\eta^\mu\eta^\nu
 - (k\cdot\eta)(k^\mu\eta^\nu+\eta^\mu k^\nu)}{(k\cdot\eta)^2-k^2}
 \right) \,.
\label{eq:D-gamma-gen}
\end{equation}
A simple inspection of which shows that it vanishes if $\mu$ or $\nu=0$.
In other words, $A_0$ photons do not propagate.  Correspondingly, their equation 
of motion becomes time independent and we can use it to remove $A_0$ from the 
effective Lagrangian (the nuclear part plus $\mathcal{L}_\mathrm{photon}$).  
We find
\begin{equation}
 \Laplace A_0 = {-}e N^\dagger \hat{Q} N \,,
\label{eq:Poisson}
\end{equation}
which is just the Poisson equation with the nucleon charge density on the 
right-hand side.  Solving this in Fourier space, where $\vNabla^2$ turns into 
the squared three-momentum, we eventually get a term
\begin{equation}
 \LL_{\text{Coulomb}}(x_0,\vx)
 = -e^2 \int\dd^3y \,
 N^\dagger(x_0,\vx)N(x_0,\vx) \,
 \frac{\eex^{-\ii\vq\cdot(\vx-\vy)}}{\vq^2}
 \left(\frac{\one + \tau^3}{2}\right)
 N^\dagger(x_0,\vy)N(x_0,\vy) \,,
\label{eq:L-Coulomb}
\end{equation}
\ie, static Coulomb potential ($\sim 1/r$ in configuration space) 
between charged nucleons.  This is a non-local interaction that really should 
be kept in the Lagrangian as a whole.  Still, to calculate Feynman diagrams it 
is convenient to split it up into a vertex
\begin{equation}
 \parbox{12em}{\includegraphics[width=13em]{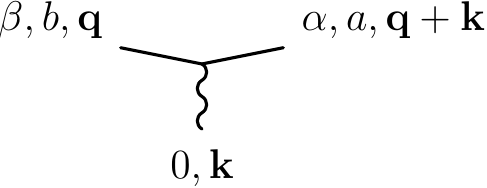}}
 \sim \; {-}{\ii e} \idxxx{\frac{\one + \tau^3}{2}}ab
 \delta\idx\alpha\beta \,,
\label{eq:Photon-Vertex-0}
\end{equation}
and a factor $\ii/\vk^2$ for each ``Coulomb-photon propagator,'' which is 
really just an expression for the static potential in momentum space.  Note 
that the sign in Eq.~\eqref{eq:Photon-Vertex-0} is arbitrary because these 
vertices really only come in pairs; we have chosen it here to coincide with 
what one would na\"ively read off from the $N^\dagger \ii D_0 N$ term.

Finding that $A_0$ photons only appear as static internal exchanges goes well in 
line that physical photons in external states have to be transverse.  Still, 
one might wonder how to treat diagrams with virtual photons coupled to a 
nuclear system (\eg, electrodisintegration), which should have $A_0$ 
contributions.  The answer is that the proper way to treat these is to add 
appropriate external currents to the Lagrangian, which then, in turn, appear on 
the right-hand side of Eq.~\eqref{eq:Poisson}.

\subsubsection{Coulomb enhancement and divergences}

\begin{figure}[htbp]
\centering
\begin{minipage}{0.25\textwidth}
 \centering
 \vcenteredhbox{\includegraphics[height=6em,clip]{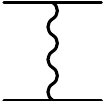}}
 \vcenteredhbox{\scalebox{1.5}{$\;\sim\;\frac{\alpha}{Q^2}$}}
\end{minipage}
\hspace{0.5em}
\begin{minipage}{0.65\textwidth}
 \centering
 \vcenteredhbox{\includegraphics[height=6em,clip]{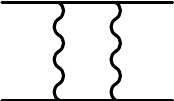}}
 \vcenteredhbox{\scalebox{1.5}{$\;\sim\;
 \frac{Q^5}{\MN}
 \left(\frac{\alpha}{Q^2}\right)^{\!2}
 \left(\frac{\MN}{Q^2}\right)^{\!2}
 = \frac{\alpha}{Q^2}\times\frac{\alpha\MN}{Q}$}}
\end{minipage}
\caption{Infrared enhancement of Coulomb-photon exchange.}
\label{fig:OneTwoPhoton}
\end{figure}

The Coulomb force is a long-range interaction: it only falls off like a power 
law ($\sim 1/r$) in configuration space.  In momentum space, it correspondingly 
has a pole at vanishing momentum transfer ($\vq^2=0$), \ie, it gives rise to an 
infrared divergence.  There are standard techniques for dealing with this, 
for example defining so-called Coulomb-modified scattering phase shifts and 
effective-range expansions well-known from quantum-mechanical scattering 
theory.  These are based on treating Coulomb effects fully nonperturbatively, 
\ie, resumming to exchange of Coulomb photons to all orders.  In the EFT power 
counting, the need for this resummation is reflected in the fact dressing a 
given two-body scattering diagram with external momenta of order $Q$ by an 
additional Coulomb photon gives a factor $\alpha\MN/Q$, \ie, an 
\emph{enhancement} if $Q \lesssim \alpha\MN$.  In Figure~\ref{fig:OneTwoPhoton} we 
show this for two-photon exchange compared to the single-photon diagram.

If a problem with Coulomb interactions is solved numerically, the 
IR divergence has to be regularized in some way to have all quantities well 
defined.  One way to do this ``screening'' the potential with a photon mass 
$\lambda$, \ie, replacing $\vq^2$ with $\vq^2 + \lambda^2$ in the 
Coulomb-photon propagator.  If this is done, $\lambda$ should be kept as small 
as possible and be extrapolated to zero for all physical observables at the end 
of the calculation.

\begin{figure}[htbp]
\centering
\includegraphics[clip,width=0.25\textwidth]{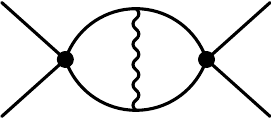}
\caption{Single photon insertion in a proton-proton bubble diagram.}
\label{fig:SinglePhotonBubble}
\end{figure}

In contrast to what one might na\"ively expect, Coulomb contributions can 
also modify the UV behavior of diagrams.  Consider, for example the insertion 
of a single photon within a bubble contributing to proton-proton scattering, as 
shown in Fig.~\ref{fig:SinglePhotonBubble}.  Power counting momenta in this 
diagram (with $C_0$ vertices) gives a factor $(Q^5/\MN)^2$ from the two loops, 
$(\MN/Q^2)^4$ from the nucleon propagators, and an obvious $\alpha/Q^2$ from 
the photon, meaning that overall we have $Q^0$, corresponding to a superficial 
logarithmic divergence of this diagram (which one can confirm with an explicit 
calculation).  This is a new feature compared to the theory with only 
short-range interactions, where this particular divergence is absent (all 
divergences are of power-law type there).  Without going into any details, we 
stress that this divergence has to be accounted for when the theory with 
Coulomb contributions is renormalized.  In particular, this example teaches us 
that in the EFT the Coulomb force is not merely an ``add-on potential'' that 
slightly shifts results, but that it has to be treated consistently along with 
the short-range interactions.

\subsubsection{Transverse photons}

Transverse photons come from the quadratic spatial derivative:
\begin{equation}
 \vec{D}^2 = (\partial^i + \ii e A^i \hat{Q})(\partial^i + \ii e A^i \hat{Q})
 = \Laplace
 + \ii e (A^i\hat{Q}\,\partial^i + \partial^i A^i \hat{Q})
 - e^2 A^2 \hat{Q} \,.
\end{equation}
We can rewrite this using
\begin{equation}
 \partial^i \left(A^i \hat{Q} N\right)
 = (\partial^i A^i) \hat{Q} N + A^i \hat{Q} \, \partial^i N \,,
\end{equation}
where the first term vanishes in Coulomb gauge.  Hence, we get the vertex
\begin{equation}
 \parbox{12em}{\includegraphics[width=13em]{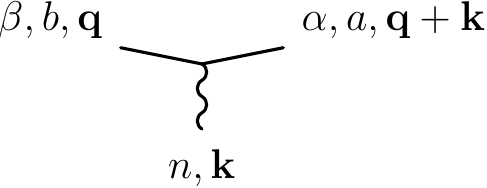}}
 \sim \; {-}\frac{e}{\MN} \idxxx{\frac{\one + \tau^3}{2}}ab
 \delta\idx\alpha\beta \, (\ii\vq)^n \,,
\label{eq:Photon-Vertex-gauge}
\end{equation}
with the momentum dependence coming from the derivative.  We leave it as an 
exercise to write down the Feynman rule for the two-photon term $\sim A^2 
\hat{Q}$.

Comparing Eq.~\eqref{eq:Photon-Vertex-gauge} with Eq.~\eqref{eq:Photon-Vertex-0}
that a diagram with the exchange of a transverse photon is suppressed by a 
factor $Q^2/\MN^2$ compared to the same topology with a Coulomb photon.
Transverse photons also have a more complicated propagator than the simple
$\ii/\vk^2$ that we found for Coulomb photons.  From Eq.~\eqref{eq:D-gamma-gen} 
we find that
\begin{equation}
 \parbox{8.25em}{\includegraphics[width=7.75em]{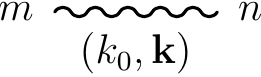}}
 \sim \; \frac{\ii}{k_0^2 - \vk^2 + \ii\eps}
 \left(\delta^{mn} - \frac{k^m k^n}{\vk^2}\right) \,.
\label{eq:Photon-Prop-gauge}
\end{equation}
which now depends on the energy $k_0$ and thus gives rise to poles in Feynman 
diagrams.  Note that the structure is somewhat different from what one 
typically sees in QED textbooks that use Lorenz/Feynman gauge.

\subsubsection{Other external currents}
\label{sec:EFT-ExtCurrents}

The covariant derivative alone only gives us photons coupled to the proton's 
charge.  However, as mentioned previously, minimal substitution only gives a 
subset of all possible electromagnetic terms.  For example, the magnetic 
coupling of photons to the nucleons (both protons and neutrons in this case) is 
given by
\begin{equation}
 \LL_\mathrm{mag} = \frac{e}{2\MN} N^\dagger
 (\kappa_0 + \kappa_1\tau^3)\,\skvec{\sigma}\cdot\vec{B}\,N \,.
\label{eq:L-mag}
\end{equation}
where $\kappa_0$ and $\kappa_1$ are the isoscalar and isovector nucleon 
magnetic moments, respectively.  That is, the low-energy constant of this 
operator has been fixed directly to the associated single-nucleon observable 
(similarly to how the nucleon mass is fixed immediately in the nonrelativistic 
theory).  With $\vec{B} = \vNabla\times\vec{A}$ and all indices written out, 
this is
\begin{equation}
 \LL_\mathrm{mag} = \frac{e}{2\MN} N^\dagger_{\alpha a}
 \idxx{\kappa_0\one + \kappa_1\tau^3}ab \idxx{\sigma^i}\alpha\beta
 \leviciv_{ijm} \partial^j A^m N^{\beta b} \,.
\label{eq:L-mag}
\end{equation}
and it gives rise to the Feynman rule
\begin{equation}
 \parbox{8em}{\includegraphics[width=8em]{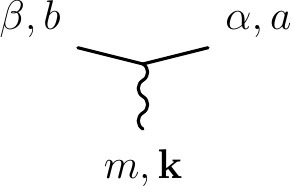}}
 \sim \; \frac{\ii e}{2\MN} \idxx{\kappa_0\one + \kappa_1\tau^3}ab \,
 \leviciv_{ijm} \idxx{\sigma^i}\alpha\beta (\ii\vk)^j \,.
\label{eq:Photon-Vertex-mag}
\end{equation}
We point out that Eq.~\eqref{eq:L-mag} only gives the leading magnetic 
coupling.  In traditional nuclear physics language, it corresponds to a 
one-body operator.  At higher orders in the EFT, there are additional 
operators, like a four-nucleon contact interaction with an additional photon.  
Such many-body terms correspond to what phenomenological approaches model as 
``meson exchange currents.''

\subsection{Example: deuteron breakup}

\begin{figure}[htbp]
\centering
\begin{minipage}{0.45\textwidth}
 \centering
 \includegraphics[height=12em,clip]{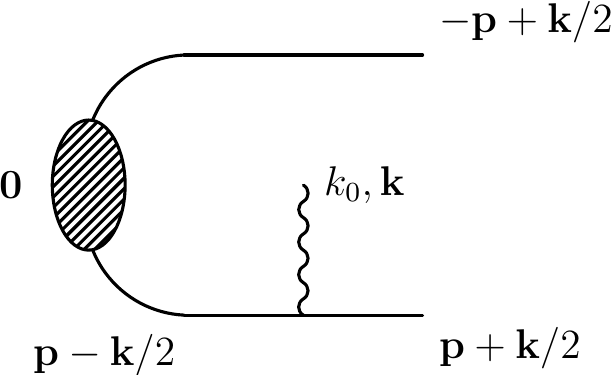}
\end{minipage}
\hspace{0.5em}
\begin{minipage}{0.45\textwidth}
 \centering
 \includegraphics[height=12em,clip]{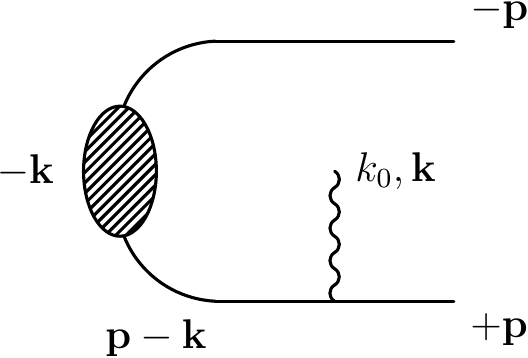}
\end{minipage}
\caption{Deuteron breakup diagram in two different kinematic frames.
 \textbf{Left:} lab frame, deuteron at rest.
 \textbf{Right:} center-of-mass frame of the outgoing nucleons.
}
\label{fig:DeutDis}
\end{figure}

As an application of the things discussed in the previous sections, we now 
consider the low-energy reaction $d\gamma \leftrightarrow np$.  By 
time-reversal symmetry, the amplitudes for the processes corresponding to the 
two possible directions of the arrow are the same.  For definiteness, we show 
the simplest diagram for the breakup reaction in Fig.~\ref{fig:DeutDis}.

The reaction can be considered in different reference frames.  In the lab 
frame (left panel of Fig.~\ref{fig:DeutDis}), the deuteron is initially at rest 
and then gets hits by a photon with four-momentum $(k_0,\vk)$.  For our 
theoretical discussion here it is more convenient to take the two outgoing 
nucleons in their center-of-mass frame, as shown in the right panel of 
Fig.~\ref{fig:DeutDis}).  To first order, we can translate between the two 
frames by boosting all nucleons lines with a momentum $\vk/2$.  A more careful 
analysis would keep track of relativistic kinematics (the external photon is 
never nonrelativistic), but this is not essential for the illustration here, so 
we can get away with interpreting $(k_0,\vk)$ to mean different things in the 
two frames.

In the $NN$ center-of-mass frame, the initial and final-state energies are
\begin{subalign}
 E_i &= k_0 + \frac{k^2}{4\MN} - \frac{\gamd^2}{\MN} \,, \\
 E_f &= \frac{p^2}{2\MN} + \frac{p^2}{2\MN} = \frac{p^2}{\MN} \,,
\end{subalign}
where we have neglected the small deuteron binding energy by setting $\Md=2\MN$.
Conservation of energy implies that
\begin{equation}
 p = \sqrt{\MN k_0 - \gamd^2 + k^2/4}
 \;\Leftrightarrow\;
 k_0 = \frac{p^2}{\MN} - \frac{k^2}{4\MN} + \frac{\gamd^2}{\MN} \,,
\end{equation}
and the energy assigned to the internal nucleon propagator has to be 
${-}\frac{k_0}{2}-\frac{\gamd^2}{2\MN}+\frac{k^2}{8\MN}$.

The blob in Fig.~\ref{fig:DeutDis} represents the deuteron vertex function 
calculated in Sec.~\ref{sec:EFT-Dibaryons}.  Taking the result found there and 
adding the spin-isospin structure, we find
\begin{equation}
\parbox{7.0em}{\centering\includegraphics[width=6.0em]{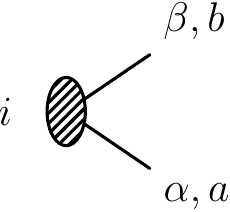}}
 \sim \; \ii\frac{1}{\sqrt8} \frac{\sqrt{8\pi\gamt}}{\MN}
 \idxx{\sigma^i\sigma^2}\alpha\beta \idxx{\tau_2}ab \,.
\label{eq:Deuteron-Vertex}
\end{equation}
With this, we can finally write down the amplitude.  For an E1 transition, the 
photon couples to the nucleon charge via the Feynman rule given in 
Eq.~\eqref{eq:Photon-Vertex-gauge}.  Combining this with the ingredients above, 
we get
\begin{multline}
 \ii\mathcal{M}_{\text{E1}} \\
 = 2 \times \frac{{-}e}{\MN}
 \idxxx{\frac{\one + \tau^3}{2}}ca
 \delta^\gamma_\alpha\,\ii({\mathbf{p}}-\vk)^j
 \dfrac{\ii}{{-}\dfrac{k_0}{2}-\dfrac{\gamd^2}{2\MN}+\dfrac{k^2}{8\MN}
 -\dfrac{({\mathbf{p}}-\vk)^2}{2\MN}+\ii\eps} \,
 \, \frac{\ii}{\sqrt8}\frac{\sqrt{8\pi\gamd}}{\MN}
 \idxx{\sigma^2\sigma^i}\beta\gamma
 \idxx{\tau_2}bc \\
 \times (\epsgammaout)^j \, (\epsdout)^i
 \, (N_{s_p})^{\alpha a} \, (N_{s_n})_{\beta b} \,.
\label{eq:np-Capture-E1-full}
\end{multline}
We have included a factor $2$ to account for the fact that in drawing the 
diagram, the photon could be coupled to either nucleon (the isospin projection 
operator ensures of course that only the proton charge gives a contribution).  
Moreover, in the second line we have included polarization vectors and spinors 
for all external particles, the spins of which we denote as $s_\gamma$, $s_d$, 
$s_p$, and $s_n$, respectively.  From the amplitude in 
Eq.~\ref{eq:np-Capture-E1-full} one can proceed to calculate the corresponding 
cross section by summing/averaging over the various initial and final states 
and integrating over the available phase space.  We skip these details and 
close by noting that the isospin part is of course completely fixed 
(``polarized'') by the experimental setup.  Hence, the spinor-isospinors are
\begin{equation}
 (N_{s_p})^{\alpha a}
 = N_{s_p}^\alpha \twodvec{1 \\ 0}^{\!a}
 \mathtext{,}
 (N_{s_n})_{\beta b}
 = N_{s_p}^\alpha \twodvec{1 \\ 0}_{\!b} \,,
\end{equation}
\ie, we can directly set $a=1$ (isospin ``up'', proton) and $b=2$ (isospin 
``down'', neutron) in the amplitude.

\begin{prob}
\emph{Exercise:} Write down the analog of Eq.~\eqref{eq:np-Capture-E1-full} 
with a magnetically-coupled (M1) photon.
\end{prob}

\subsection{Chiral effective field theory}
\label{sec:EFT-Chiral}

Pionless EFT provides a simple yet powerful framework to describe few-nucleon 
systems at very low energies, but its name implies its natural limitation, \ie, 
the inability to describe physics at energy scales where pion exchange can be 
resolved.  Certainly this becomes important for scattering calculations at 
momentum scales larger than the pion mass.  But nuclear binding also 
generally increases with increasing number $A$ of bound nucleons, which 
translates to larger typical momentum scales within the nucleus.  There are 
indications that pionless EFT still converges for $A=4$, but the question is 
not fully settled.

The construction of an effective field theory of nucleons and pions was 
pioneered by Weinberg in the early 1990s~\cite{Weinberg:1990rz,Weinberg:1991um},
proposing a scheme to construct a nuclear potential based on Feynman diagrams 
from chiral perturbation theory.  This theory, which is constructed as an 
expansion around the so-called ``chiral limit,'' in which the pions are exactly 
massless Goldstone bosons.  The resulting theory, which has been applied with 
great success in the purely pionic and single-nucleon sector, treats the pion 
mass as a \emph{small} scale and thus has a power counting designed for typical 
momenta $Q\sim\mpi$.

For two or more nucleons, the theory is highly nonperturbative, which motivated 
Weinberg to develop a scheme where the power counting is applied to the 
\emph{potential} instead of the amplitude, as we have otherwise done throughout 
this section.  Kaplan~\etal~\cite{Kaplan:1996xu,Kaplan:1998we,Kaplan:1998tg} 
proposed a different scheme where pions are included perturbatively on top of a 
leading order given by pionless EFT.  This approach has, however, been found not 
to converge in channels where pion exchange generates a singular attractive 
interaction~\cite{Fleming:1999ee}.  It is thus understood today that pions in
general have to be treated nonperturbatively, in a framework generally referred 
to as ``chiral effective field theory.''  How exactly this should be 
implemented, however, is still a matter of debate.  Instead of summarizing this 
here, we refer the reader to the 
literature (see, e.g., Refs.~\cite{Beane:2000fx,Bedaque:2002mn,Nogga:2005hy,
Birse:2005um,Epelbaum:2008ga,Machleidt:2011zz,Epelbaum:2006pt,Hammer:2012id,
Long:2016vnq}).

\subsubsection{Leading-order pion-nucleon Lagrangian}

For a thorough introduction to the field of chiral perturbation theory, we 
recommend the reader to study the lecture notes of Scherer and 
Schindler~\cite{Scherer:2012xha} as well as the vast original literature cited 
therein.  It uses an elaborate formalism to construct the most general 
pion-nucleon Lagrangian that is invariant under chiral transformations 
(individual rotations of left- and right-handed nucleon fields) plus terms 
implementing the explicit breaking of chiral symmetry due to finite quark 
masses.  In the conventions of Ref.~\cite{Scherer:2012xha}, the leading-order 
pion-nucleon Lagrangian is
\begin{equation}
 \LL_{\pi N}^{(1)} + \LL_2^\pi
 = \bar{\psi}\left(\ii\slashed{D} - \MN
 +\frac{g_A}{2}\gamma^\mu\gamma^5 u_\mu\right)\psi
 + \frac{\fpi^2}{4}\Tr\left[(\partial^\mu U)^\dagger (\partial_\mu U)
 + \chi U^\dagger + U \chi^\dagger\right] \,.
\label{eq:L-pi-N-chiral}
\end{equation}
Here, $\psi$ is the nucleon Dirac field, and the matrix-valued field
\begin{equation}
 U \equiv = \exp\!\left(\ii\frac{\skvec{\tau}\cdot\skvec{\pi}}{\fpi}\right)
\label{eq:U-pi}
\end{equation}
collects the pion fields in an exponential representation.  $D_\mu$ here is the 
so-called chiral covariant derivative that couples the pion field to the 
nucleons.  The matrix
\begin{equation}
 \chi = 2 B_0 \, \diag(m_q,m_q) \,,
\end{equation}
where $m_q$ is the light quark mass (in the exact isospin limit, $m_u=m_d=m_q$) 
contains the effect from explicit chiral symmetry breaking.  Via the 
Gell-Mann--Oakes--Renner relation,
\begin{equation}
 \mpi^2 = 2B_0 m_q \,,
\label{eq:GMOR}
\end{equation}
one can show that $\chi$ generates a mass term for the pion field.

\begin{prob}
\emph{Exercise:} Expand the exponential in Eq.~\eqref{eq:U-pi} to show that 
$\frac{\fpi^2}{4} \Tr\left[(\partial^\mu U)^\dagger (\partial_\mu U)\right]$ 
generates a kinetic term for the pion field $\skvec{\pi}$ plus higher-order 
pion self interactions.
\end{prob}

After a couple of steps, which we encourage the reader to follow 
based on the definitions given in Ref.~\cite{Scherer:2012xha}, the leading 
terms in the Lagrangian are found to be
\begin{multline}
 \LL_{\pi N}^{(1)} + \LL_2^\pi
 = \bar{\psi}\left(\ii\slashed{\partial}-\MN\right)\psi
 + \frac{1}{2}(\partial^\mu\skvec{\pi})\cdot(\partial_\mu\skvec{\pi})
 - \frac{1}{2}\mpi^2 \skvec{\pi}^2\\
 - \frac{g_A}{2\fpi}\bar{\psi}\gamma^\mu\gamma^5
 (\skvec{\tau}\cdot\partial_\mu\skvec{\pi})\psi
 + \frac{\ii}{8\fpi^2}\bar{\psi}\gamma^\mu
 (\skvec{\tau}\cdot\skvec{\pi})(\skvec{\tau}\cdot\partial_\mu\skvec{\pi})\psi
 + \cdots \,.
\label{eq:L-pi-N-chiral-LO}
\end{multline}
Comparing this to our pseudoscalar model~\eqref{eq:L-pi-N-PS} in 
Sec.~\ref{sec:EFT-NonRelFerm}, we see that the pion-nucleon coupling now comes 
with explicit derivatives, as required by chiral symmetry.  After a 
Foldy-Wouthuysen transformation, however, one-pion-exchange in the 
nonrelativistic limit gives the same structure as in 
Eq.~\eqref{eq:pi-N-EOM-NR}.  There is no explicit $\sigma$ field in 
Eq.~\eqref{eq:L-pi-N-chiral-LO}; in the chiral theory, this particle only 
appears as a resonance generated by two-pion exchange.

\begin{acknowledgement}
We thank Dick Furnstahl and Bira van Kolck and for various stimulating 
discussions.  Moreover, SK is grateful to Martin Hoferichter for the insights 
into Coulomb-gauge quantization presented in Sec.~\ref{sec:EFT-EM}.
\end{acknowledgement}

\end{document}